\documentclass[apj]{emulateapj}

\usepackage[utf8]{inputenc}
\usepackage{bold-extra}
\usepackage{graphicx}
\usepackage{amsmath,amssymb}
\usepackage{mathptmx}
\usepackage[english]{babel}
\usepackage{booktabs}
\usepackage{multirow}
\usepackage{natbib}
\usepackage[breaklinks,colorlinks,urlcolor=blue,citecolor=blue,linkcolor=blue]{hyperref}
\usepackage{color}
\usepackage{verbatim}
\usepackage{rotating}

\newcommand{\beq}	{\begin{equation}}
\newcommand{\eeq}	{\end{equation}}
\newcommand{\beqs}	{\begin{displaymath}}
\newcommand{\eeqs}	{\end{displaymath}}
\newcommand{\beqa}  {\begin{eqnarray}}
\newcommand{\eeqa}  {\end{eqnarray}}
\newcommand{\beqas}	{\begin{eqnarray*}}
\newcommand{\eeqas}	{\end{eqnarray*}}
\newcommand{\cc}	{\centering}

\def\kb{\ifmmode {k_{\rm B}} \else $k_{\rm B}$\fi}
\def\mp{\ifmmode {m_{\rm p}} \else $m_{\rm p}$\fi}
\def\mbar{\ifmmode {\bar{m}} \else $\bar{m}$\fi}
\def\msun{\ifmmode {\rm M_{ \odot}} \else ${\rm M_{\odot}}$\fi}
\def\msuny{\ifmmode {\rm  M_{\odot}~{\rm yr^{-1}}} \else ${\rm M_{\odot}~{\rm yr^{-1}}}$\fi}
\def\lstar{\ifmmode {L^{*}} \else $L^{*}$\fi}
\def\cmv{\ifmmode {\rm cm^{-3}} \else ${\rm cm^{-3}}$\fi}
\def\cmc{\ifmmode {\rm cm^{-2}} \else ${\rm cm^{-2}}$\fi}
\def\kms{\ifmmode {\rm km\:s^{-1}} \else $\rm km\:s^{-1}$\fi}

\def\r200{\ifmmode {r_{\rm 200}} \else $r_{\rm 200}$\fi}
\def\rvir{\ifmmode {R_{\rm vir}} \else $R_{\rm vir}$\fi}
\def\mvir{\ifmmode {M_{\rm vir}} \else $M_{\rm vir}$\fi}
\def\rcgm{\ifmmode {r_{\rm CGM}} \else $r_{\rm CGM}$\fi}
\def\cgm2{\ifmmode {{\rm CGM^2}} \else ${\rm CGM^2}$\fi}
\def\nh{\ifmmode {n_{\rm H}} \else $n_{\rm H}$\fi}

\def\tinf{\ifmmode {t_{\rm infall} } \else $t_{\rm infall}$\fi}
\def\tdyn{\ifmmode {t_{\rm dyn}} \else $t_{\rm dyn}$\fi}
\def\tcool{\ifmmode {t_{\rm cool}} \else $t_{\rm cool}$\fi}
\def\vturb{\ifmmode {\sigma_{\rm turb}} \else $\sigma_{\rm turb}$\fi}

\def\ah{\ifmmode {\alpha_{\rm hot}} \else $\alpha_{\rm hot}$\fi}
\def\mhot{\ifmmode {M_{\rm hot}} \else $M_{\rm hot}$\fi}
\def\th{\ifmmode {T_{\rm hot}} \else $T_{\rm hot}$\fi}
\def\nhot{\ifmmode {n_{\rm hot}} \else $n_{\rm hot}$\fi}

\def\ac{\ifmmode {\alpha_{\rm cool}} \else $\alpha_{\rm cool}$\fi}
\def\tc{\ifmmode {T_{\rm cool}} \else $T_{\rm cool}$\fi}
\def\ncool{\ifmmode {n_{\rm cool}} \else $n_{\rm cool}$\fi}
\def\mcool{\ifmmode {M_{\rm cool}} \else $M_{\rm cool}$\fi}
\def\dmcool{\ifmmode {\dot{M}_{\rm cool}} \else $\dot{M}_{\rm cool}$\fi}

\def\nint{\ifmmode {n_{\rm IT}} \else $n_{\rm IT}$\fi}
\def\mint{\ifmmode {M_{\rm IT}} \else $M_{\rm IT}$\fi}
\def\fvint{\ifmmode {f_{\rm V,IT}} \else $f_{\rm V,IT}$\fi}

\def\ncz{\ifmmode {n_{\rm 0,c}} \else $n_{\rm 0,c}$\fi}
\def\anc{\ifmmode {a_{\rm n,c}} \else $a_{\rm n,c}$\fi}
\def\fvc{\ifmmode {f_{\rm V,cool}} \else $f_{\rm V,cool}$\fi}
\def\fac{\ifmmode {f_{\rm c}} \else $f_{\rm c}$\fi}
\def\uhm{\ifmmode {\left< U \right>_{H,M}} \else $\left< U \right>_{H,M}$\fi}
\def\umean{\ifmmode {\left< U \right>}\else $\left< U \right>$\fi}
\def\ncoolm{\ifmmode {\left< n_{\rm cool} \right>_{M}} \else $\left< n_{\rm cool} \right>_{M}$\fi}

\def\rcl{\ifmmode {R_{\rm cl}} \else $R_{\rm cl}$\fi}
\def\vcl{\ifmmode {v_{\rm cl}} \else $v_{\rm cl}$\fi}
\def\ncl{\ifmmode {n_{\rm cl}} \else $n_{\rm cl}$\fi}
\def\nls{\ifmmode {N_{\rm cl}} \else $N_{\rm cl}$\fi}
\def\Mcl{\ifmmode {M_{\rm cl}} \else $M_{\rm cl}$\fi}

\newcommand{\HI}{\rm H{\textsc{~i}}}
\newcommand{\CII}{\rm C{\textsc{~ii}}}
\newcommand{\CIII}{\rm C{\textsc{~iii}}}
\newcommand{\CIV}{\rm C{\textsc{~iv}}}
\newcommand{\SiII}{\rm Si{\textsc{~ii}}}
\newcommand{\SiIII}{\rm Si{\textsc{~iii}}}
\newcommand{\SiIV}{\rm Si{\textsc{~iv}}}
\newcommand{\MgII}{\rm Mg{\textsc{~ii}}}
\newcommand{\OVI}{\rm O{\textsc{~vi}}}
\newcommand{\OVII}{\rm O{\textsc{~vii}}}
\newcommand{\OVIII}{\rm O{\textsc{~viii}}}
\newcommand{\NV}{\rm N{\textsc{~v}}}

\definecolor{mgreen}{rgb}{0, 0.5, 0}

\begin{document}

\defcitealias{FSM17}{FSM17}
\defcitealias{FSM20}{FSM20}
\defcitealias{F22}{F22}
\defcitealias{Werk14}{W14}
\defcitealias{Werk13}{W13}
\defcitealias{Prochaska17}{P17}
\defcitealias{Klypin02}{KZS02}
\defcitealias{HM12}{HM12}
\defcitealias{KS19}{KS19}

\title{The cool circumgalactic medium of low-redshift star-forming galaxies: \\ I - Empirical model and mean properties}

\author{
Yakov Faerman \altaffilmark{1 *},
Jessica K. Werk \altaffilmark{1},
}
\shorttitle{Modeling the Cool CGM}

\altaffiltext{*}{e-mail: \href{mailto:yakov.faerman@gmail.com}{yakov.faerman@gmail.com}}
\altaffiltext{1}
{Astronomy Department, University of Washington, Seattle, WA 98195, USA}

\begin{abstract}
We present an analytic model for the cool, $T \approx 10^4$~K, circumgalactic medium (CGM), describing the gas distribution, thermal and ionization state. Our model assumes (total) pressure equilibrium with the ambient warm/hot CGM, photoionization by the metagalactic radiation field, and allows for non-thermal pressure support, parametrized by the ratio of thermal pressures, $\eta = P_{\rm hot,th}/P_{\rm cool,th}$. We apply the model to the COS-Halos data set and find that a nominal model with $\eta = 3$, gas distribution out to $r \approx 0.6 \rvir$, and $\mcool = 3 \times 10^9$~\msun, corresponding to a volume filling fraction of $\fvc \approx 1\%$, reproduces the mean measured column densities of \HI~and low/intermediate metal ions (\CII, \CIII, \SiII, \SiIII, \MgII). Variation of $\pm 0.5$~dex in the non-thermal pressure or gas mass encompasses $\sim 2/3$ of the scatter between objects. Our nominal model underproduces the measured \CIV~and \SiIV~columns, and we show these can be reproduced with (i) a cool phase with $\mcool \approx 10^{10}$~\msun~and $\eta \approx 5$, or (ii) an additional component at intermediate temperatures, of cooling or mixing gas, with $M \approx 1.5 \times 10^{10}$~\msun~and occupying $\sim 1/2$ of the total CGM volume. For cool gas with $\fvc \approx1\%$ we provide an upper limit on the cloud sizes, $\rcl \lesssim 0.5$~kpc. Our results suggest that for the average galaxy CGM, the mass and non-thermal support in the cool phase are lower than estimated in previous works, and extreme scenarios for galactic feedback and non-thermal support may not be necessary. We estimate the rates of cool gas depletion and replenishment, and find accretion onto the galaxy can be entirely offset by condensation, outflows, and IGM accretion, allowing $\dot{M}_{\rm cool}\sim 0$ over long timescales.
\end{abstract}



\section{Introduction}
\label{sec:intro}

Absorption observations of the low-redshift CGM reveal massive, metal-enriched, and highly multiphase halos, with ions from \MgII~to \OVIII, tracing temperatures from $\sim 10^4$~K to $\sim 10^6$~K \citep{TPW17}. The physical conditions in the CGM are set by the same processes that affect the evolution of galaxies and determine their morphologies, including accretion from the IGM into the halo and onto the galaxy, gas stripping from satellite galaxies, and feedback from the galaxy, in the form of winds and radiation from supernovae and active galactic nuclei (AGN).

The warm/hot CGM has been studied by many recent works presenting different physical models and exploring various assumptions and scenarios \citep{FSM17,MP17,Stern18,Qu18a,Qu18b,MW18,Voit19,Stern19,FSM20}. These models, addressing the gas spatial distribution, kinematics, thermal properties, and ionization states, enable relating the CGM observables to the gas physical properties and the processes that shape the CGM and the galaxy. They also produce different gas distributions that may be tested with future observations in the radio (SZ, FRBs), and UV/X-ray absorption and emission (see Singh~et~al., in prep., for a comparison between models).

The cool component of the CGM, at $T \sim 10^4$~K, is not less interesting, as it may dominate the accretion from the CGM onto the galaxy \citep{Opp18c}, replenishing the ISM and providing fuel for star-formation and SMBH growth \citep{Putman12,Somer15}.

Recent decades have seen significant progress in observational studies of the cool CGM. At low redshift, this phase is observed mainly in UV absorption, using COS/HST (\citealp{Prochaska11,Werk13,Tumlinson13}, but see \citealp{PSFM22} for emission predictions). Measurements include both hydrogen transitions and metal lines, allowing estimation of the gas metallicity \citep{Prochaska17,Wotta19,Berg23} and showing that it is significantly metal-enriched. At higher redshifts, cool gas is observed both from space and the ground, in absorption \citep{HP13,Rudie13,Lehner14} and emission \citep{Cantalupo14,Borisova16,Cai18}. Neutral and molecular gas at temperatures below $10^4$~K is also detected \citep{Neeleman15,Neeleman16}.

These observations motivated many theoretical studies, with different approaches, ranging from analytic estimates for the gas formation and evolution timescales \citep{MB04}, phenomenological modeling \citep{Stern16,Haislmaier21,Sameer21}, toy models addressing different scenarios for the gas kinematics \citep{LM19,Afruni19}, and numerical simulations of gas physics, including the effects of radiative processes, turbulence, magnetic fields, and cosmic rays \citep{McCourt12,Arm17,Ji19,Butsky20,Sparre20,Gronke22}. 

Larger, zoom-in and cosmological simulations are crucial for capturing the effects of the cosmic web and galaxy evolution on the structure and thermodynamics of the CGM \citep{Nelson18b,Opp18b,RF19,Hafen19a,Appleby21}. However, these numerical experiments face several challenges when used to study the CGM. First, the gas properties vary significantly with the subgrid recipes employed to model stellar and black hole feedback \citep{Kelly22}, tuned to reproducing the observed galactic stellar populations. Second, the spatial resolution in the CGM is often limited. Recent studies that focused more attention on the CGM by increasing the spatial resolution (\citealp{Hummels19,Peeples19,Voort19}, and see also \citealp{Nelson20}) showed that while the properties of the volume-filling warm/hot gas may be converged, this is not the case for the cool CGM properties and observables. Finally, the effects of small-scale physical processes are often not included in these simulations and may be especially significant for the cool gas (\citealp{Hopkins20}).

Given these challenges, it is interesting to employ analytic models to study the cool phase of the CGM. These can test a variety of physical scenarios and processes, and be applied to observations to address interesting questions that remain open about this phase -- what is its spatial distribution, thermal properties, and morphology? What are the physical processes that set its properties? How does it evolve and interact with the galaxy? 

In this work we present a physically-motivated phenomenological model for the cool CGM in combination with the detailed warm/hot CGM model developed in \citetalias{FSM20}. We address the spatial distribution of the cool gas and its ionization state, including the effects of photoionization and considering non-thermal support. We apply this model to the COS-Halos measurements of HI and metal ions to estimate the cool gas extent, density, and mass in an average star-forming galaxy. This is the first study performing forward-modeling of the observed metal ion column densities in the cool CGM within the context of a physically-motivated model for the warm/hot gas.

This manuscript is structured as follows. We discuss the observational data we model and our goals in \S\ref{sec:goals}, describe our model setup in \S\ref{sec:setup}, and present our main results for the column densities of cool gas in an average galaxy in \S\ref{sec:models}. We then show additional variations of the model in \S\ref{sec:advanced} and provide tools for inferring gas properties in individual sightlines. We address some uncertainties of our model in \S\ref{sec:uncertain}, discuss the morphology and the depletion-replenishment cycle of cool CGM in \S\ref{sec:disc}, and summarize in \S\ref{sec:summary}.

\section{Goals and Observational Data}
\label{sec:goals}

In this work we model the hydrogen and metal ion column densities in the CGM of star-forming galaxies at low redshifts, measured as part of the COS-Halos survey. We now briefly describe the survey, the data, and the goals of our study.

The COS-Halos survey used the Cosmic Origins Spectrograph (COS) aboard the {\emph{HST}} to study gas around \lstar~galaxies at redshifts $0.1 \lesssim z \lesssim 0.4$ \citep{Tumlinson11,Werk12}. The survey, carried out during HST Cycle 17, the first cycle in which COS was available, remains the most complete CGM survey of $L^*$ galaxies at $z < 1$ probing the inner CGM, at impact parameters of up to $\sim 0.6$~\rvir~\citep{Werk13}. It revealed that the CGM is extended, massive, and highly multiphase, with ions ranging from low (\MgII, \CII, \SiII), through intermediate species (\CIII, \SiIII), and up to even more ionized gas, traced by higher ions (\NV~and \OVI)\footnote{~Other studies and data sets at low-z span different halo masses, galaxy morphologies, and environments \citep{Stocke13,Bordoloi14,Borthakur15,Johnson15,Keeney17,Burchett19,Chen20,Zahedy21}. Data from the \cgm2~survey, for example, extend to larger impact parameters, beyond $\rvir$~\citep{Wilde21,KT22}}. The survey also measured the galaxy properties, including stellar masses and star formation rates. Motivated by the measurements of high \OVI~column densities in star-forming galaxies (sSFR$>10^{-11}~{\rm yr^{-1}}$), \citetalias{FSM17} and \citetalias{FSM20} constructed models for the warm/hot CGM, which successfully reproduced the observed \OVI~column density profile, as well as the \OVII~and \OVIII~column densities measured in the MW in the X-ray \citep{Bregman07,Gupta12,Fang15}.

In this work, we extend the \citetalias{FSM20} framework by adding a cool, photoionized component to the CGM. We then apply this model to the COS-Halos measurements of \HI~and the low/intermediate metal ions to infer the model parameters consistent with observations and constrain the gas properties, such as density and mass. For this purpose, we use the same sub-sample of star-forming galaxies addressed by \citetalias{FSM20}. The column densities are from \citealp{Werk13} and \citealp{Prochaska17}. For the metal ions, we focus on the \CII-\CIV, \SiII-\SiIV~and \MgII~columns. These ions span a range of ionization potentials and probe different gas densities~\footnote{~Outputs for similar ions of other elements (nitrogen, sulfur, and iron, for example) are calculated and are not shown for compactness of presentation.}. The columns modeled in this work are the total column densities measured for each sightline, summed over all detected sub-components. We plot the measured columns and their errors in Figures~\ref{fig:hi}-\ref{fig:mg2col}, and provide the underlying data in Table~\ref{tab:obscol} in Appendix~A.

One notable feature of the low ion measurements is the scatter in column densities between different galaxies at a given impact parameter. For example, the \HI~columns (Figure~\ref{fig:hi}) have a scatter of $>1$~dex at impact parameters $<0.4$~\rvir. The metal ions (Figures~\ref{fig:carbon}-\ref{fig:mg2col}) can have both upper and lower limits, differing by $0.5-1.0$~dex, at similar (projected) distances from the galaxy. This is in contrast to the \OVI~column densities measured for these galaxies \citep{Tumlinson11}, which show a much smaller scatter between different objects in the sample (see Fig.~10 in \citetalias{FSM20}).

Our main goal in this work is to reproduce the typical, mean column densities of \HI~and metal ions, and their scatter. This is a first order approximation, addressing the properties of the cool gas in an average star-forming galaxy\footnote{~The columns of individual objects, which may be related to the galaxy properties and histories, will be the focus of a separate study.}. For easier comparison to our models, we bin the individual measurements and the binned data are plotted as large black and grey markers, with the errors bars representing the $1-\sigma$ scatter in each bin. For the purpose of binning, we treat upper and lower limits as measurements, unless all the data in a bin are upper or lower limits. The vertical thin dashed lines in the left panel of each figure show the boundaries of the bins.

We note a few other points. First, while the small scale morphology and kinematics of the cool gas are beyond the scope of this work, in \S\ref{subsec:disc_sizes} we show that our model provides upper limits on the gas cloud size. Second, our model is agnostic to the origin or formation channel of the cool gas, and we discuss these briefly in \S\ref{subsec:disc_res}. With these goals and reservations in mind, we now present our model setup for the cool gas.

\section{Model Setup}
\label{sec:setup}

Two important inferences have been made from modeling the COS-Halos absorption measurements: (i) the cool gas densities are lower than those expected from thermal pressure equilibrium with the hot phase (\citealp[hereafter W14]{Werk14}), and (ii) the cool gas mass, traced by \HI, is $\mcool \gtrsim 10^{10} ~\msun$, a significant fraction of the total baryons associated with these halos (\citetalias{Werk14}, \citealp[hereafter P17]{Prochaska17}). We revisit these two points in this work and aim to quantify what gas densities and masses are required to reproduce the measured absorption. To do this, we add a cool gas phase to the \citetalias{FSM20} model for the warm/hot CGM, and we now describe the parameters we choose for constructing the cool gas distribution.

First, we address non-thermal pressure support, setting the gas density. In \citetalias{FSM20}, the amount of non-thermal pressure in the warm/hot phase is given by the parameter $\alpha_{\rm tot} = P_{\rm hot,tot}/P_{\rm hot,th} \equiv \ah$ (see also Eq.~2~in~\citetalias{F22}). In this work, we allow the cool phase to have a separate non-thermal pressure component, $\ac = P_{\rm cool,tot}/P_{\rm cool,th}$, and define the ratio
\beq\label{eq:eta}
\eta \equiv \ac/\ah ~~~
\eeq
The high detection rate of cool gas in the CGM suggests that it is abundant and long-lived, rather than a transient phase. Motivated by this, we assume local total pressure equilibrium between the phases, $P_{\rm hot,tot}(r) = P_{\rm cool,tot}(r) \equiv P_{\rm tot}(r)$, giving 
\beq\label{eq:eta2}
\eta = P_{\rm hot,th}/P_{\rm cool,th} ~~~
\eeq
For a given distribution of warm/hot gas, setting $\eta$ allows us to write the cool gas density at a given distance $r$ from the center of the halo (i.e. from the galaxy)
\beq\label{eq:ncool}
\ncool(r) = \nhot \frac{\ah}{\ac} \frac{T_{\rm hot}}{\tc} = \frac{P_{\rm hot,th}/\kb}{\eta \tc} ~~~,
\eeq
where $P_{\rm hot,th}/\kb=\nhot \th$, and $\tc$ is the temperature of the cool phase, set by heating/cooling equilibrium with the metagalactic radiation field (MGRF)\footnote{~In this work we assume the cool gas heating and photoionization are dominated by the MGRF. We discuss the effect of different MGRF models and a possible contribution of galactic radiation in \S\ref{subsec:disc_rad}.}, and it depends on the gas density and metallicity. To obtain $\ncool$, we assume the metallicity at a given radius and solve for the density and temperature iteratively. We start by inserting $\tc = 10^4$~K into Eq.~\eqref{eq:ncool} and solve for the density. We then use Cloudy 17.00 \citep{Ferland17} to obtain the equilibrium temperature for this density, adopting the \citealp{HM12} (hereafter \citetalias{HM12}) MGRF, and repeat the calculation until the temperature converges. For the densities estimated for the cool CGM, $10^{-1} < \ncool/\cmv <10^{-5}$ (\citetalias{Werk14},\citetalias{Prochaska17}) and at a constant metallicity, $\tc$ varies weakly with $n$\footnote{At $\ncool < 10^{-5}$~\cmv, the gas equilibrium temperature has a stronger variation with density. As we see in the next section, the cool gas densities in our models are above this limit.}. We show $\tc$ as a function of radius for different values of $\eta$ in the right panel of Figure~\ref{fig:thermal}.

Given the weak variation of temperature with radius, the density profile of the cool gas is set mainly by $\eta$ and by the pressure in the warm/hot phase. The non-thermal support in the warm/hot phase varies with radius, and for the fiducial FSM20 model it is in the range $\ah \approx 1.5-3$. In this work, we use $\eta$, rather than $\ac$, as one of the parameters for our model of the cool gas, and assume $\eta$ to be constant with radius, for simplicity. For $\eta=1$, pressure equilibrium between the two phases implies thermal pressure equilibrium, and the density ratio is given by the inverse of the temperature ratio. $\eta>1$ lowers the density in the cool gas, and it is motivated empirically by the analysis in \citetalias{Werk14} (see their Section~5.3 and Fig.~12). Theoretically, higher non-thermal support in the cool phase is expected, for example, for magnetic fields if the cool gas forms from the hot component (see \citealp{Nelson20}) and the magnetic flux is frozen into the condensing clouds, or other mechanisms that couple more strongly to denser gas.

Second, we define the spatial distribution of the cool gas by setting its volume filling fraction, 
\beq\label{eq:fvcool}
\fvc \equiv dV_{\rm cool}/dV ~~~,
\eeq
to be non-zero in a radial range $r_1<r<r_2$, where $r_1=0.05 \rvir$ and $r_2<\rcgm$. The inner boundary is chosen to exclude the region close to the galaxy, where winds/outflows and interaction with the galaxy may be dominant, requiring a more detailed model (see \citealp{Fielding22}, for example). As shown in Figures~\ref{fig:hi}-\ref{fig:mg2col}, the observations we model in this work probe impact parameters up to $\sim 0.6 \rvir$.

The volume filling fraction, together with the gas density, set the mass distribution and the total mass in the cool phase:
\beq
\begin{split}
M_{\rm cool} & = \int_{0}^{\rcgm}{\rho_{\rm cool}(r) dV_{cool}} \\
       & = 4 \pi \int_{r_1}^{r_2}{\fvc(r) \mbar(r) \ncool(r) r^2 dr} ~~~,
\end{split}
\label{eq:mcool1}
\eeq
where $\mbar$ is the mean mass per particle, calculated as a function of radius given the gas ionization state. In this work we assume a mass fraction of $Y=0.25$ in helium, and $\mu \equiv \mbar/\mp$ ranges from $0.59$~in fully ionized gas to $1.23$ in neutral gas.

To compare our models with observations, we calculate the ion column densities as functions of the impact parameter. To do this, we use Cloudy to calculate the gas ionization state (i.e. ion fractions) assuming photoionization equilibrium in optically thin gas in the presence of the HM12 MGRF, for a range of gas densities and temperatures. The \HI~column as measured by an external observer through the CGM at an impact parameter $h$ from the galaxy, is then given by
\beq \label{eq:hi_col}
N_{\HI}(h) = \int_{s}{n_{\rm cool}(r) \fvc(r) f_{\HI}(r)ds} ~~~,
\eeq
where $s = \sqrt{h^2-r^2}$ is the coordinate along the line of sight. For an assumed (relative to solar) metallicity profile $Z'(r)$, the column for an ion of element $X$ with abundance $A_{\rm X}$ is
\beq \label{eq:met_col}
N_{ion}(h) = \int_{s}{n_{\rm cool}(r) \fvc(r) A_{\rm X} Z'(r) f_{\rm ion}(r) ds} ~~~.
\eeq
In this work we use the \citealp{Asplund09} values for the solar metal abundances, with $A_{\rm C} = 2.7 \times 10^{-4}$, $A_{\rm Si}=3.2 \times 10^{-5}$, $A_{\rm Mg}=4.0 \times 10^{-5}$, for carbon, silicon, and magnesium, respectively. Next we apply this model to the data.

\section{Model Demonstration - Basic}
\label{sec:models}

In this section we present three sets of parameter combinations, in which we vary the gas physical properties, examine how these variations affect the gas observables - the \HI~and metal ions column densities - and how observations constrain the properties of the cool CGM.

First, we examine the amount of non-thermal pressure in the cool phase, set by $\eta$ ~(see Eq.~\ref{eq:eta}), and show how it affects the gas ionization state (Pressure Set, $\#1$, \S\ref{subsec:mod_pressure}). Second, we address the gas radial distribution and vary $r_2$, the maximal radius at which cool gas is present (Radius Set, $\#2$, \S\ref{subsec:mod_radius}). To isolate the effect of these parameters on the column densities, we set the gas mass in these two sets to $\mcool = 3.0 \times 10^{9}$~\msun~by adjusting the value of \fvc, the gas volume filling fraction, between models (see Eq. \ref{eq:mcool1}). Finally, we vary $\mcool$, asking what masses are needed to reproduce the measured columns (Mass Set, $\#3$, \S\ref{subsec:mod_mass}).

As described in \S\ref{sec:setup}, we set $\eta$, $\fvc$, and the gas metallicity, $Z'$, to be constant as functions of radius\footnote{~We do this to make the model as simple as possible, allowing better understanding of our results. This assumption can be relaxed in future work, to examine if variation with radius is required by the data. In \S\ref{subsec:adv_estim} we show how the measured column densities can be used to constrain these parameters for individual objects and lines of sight.}. We adopt a value of $Z'= 0.3$~solar, motivated by the median metallicity inferred for the COS-Halos absorbers by \citetalias{Prochaska17}\footnote{~We note this is different from the metallicity profile of the warm/hot gas, which varies with radius between $Z'=1.0$ near the galaxy and $Z'=0.3$~at~\rcgm~(see Fig.~3 in \citetalias{FSM20}).} (see also \citealp{Wotta16,Wotta19}). We adopt the fiducial \citetalias{FSM20} parameters for the warm/hot gas distribution, and discuss how variation in the ambient pressure profile affects our results in \S\ref{subsec:disc_sens}. To calculate the ion fractions, we use the \citetalias{HM12} field at $z=0.2$, the median redshift of the galaxies in our sample, and in \S\ref{subsec:disc_rad} we address a possible contribution from galactic radiation, and the redshift evolution of the MGRF ionizing flux.

\begin{figure*}[t]
\includegraphics[width=0.99\textwidth]{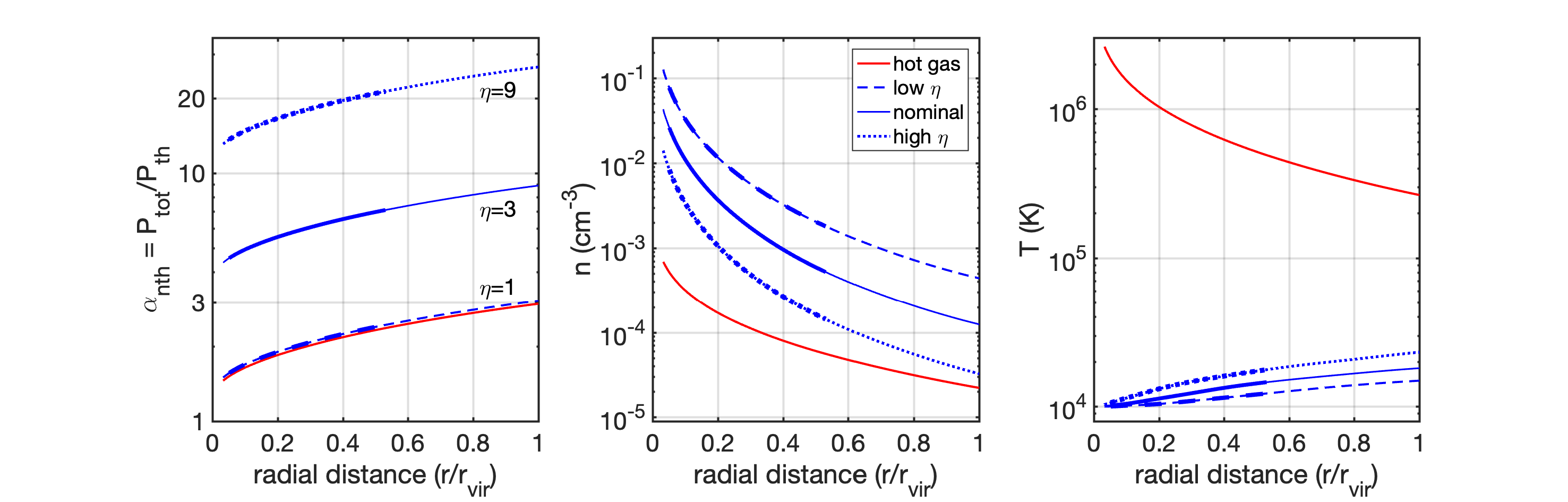}
 \caption{Gas thermal properties - non-thermal support (left), density (middle) and temperature (right). The red curves show the properties of the warm/hot gas in the \citetalias{FSM20} fiducial model, and the blue curves are for the cool phase described in this work. The solid curves show the nominal model, and the models in the Radius and Mass sets ($\#2$ and $\#3$) have the same profiles. The dashed (dotted) curves show models with low (high) non-thermal support (Pressure set, $\#1$). The thick part of each curve shows the gas nominal spatial extent, $r_2 = 0.55\rvir$. {\bf Left:} $\eta \equiv \ac/\ah$ is constant with radius, and the actual amount of non-thermal support, given by $\ac$, follows the non-thermal support in the warm/hot gas. {\bf Middle:} For a given total pressure profile set by the hot gas, higher non-thermal support in the cool phase corresponds to lower volume density. {\bf Right:} The gas temperature, set by heating/cooling equilibrium with the MGRF, depends on the gas density and varies with radius and, to a smaller extent, between models (see \S\ref{subsec:mod_pressure} for details).}
 \label{fig:thermal}
\end{figure*}

\begin{figure*}[t]
\includegraphics[width=0.99\textwidth]{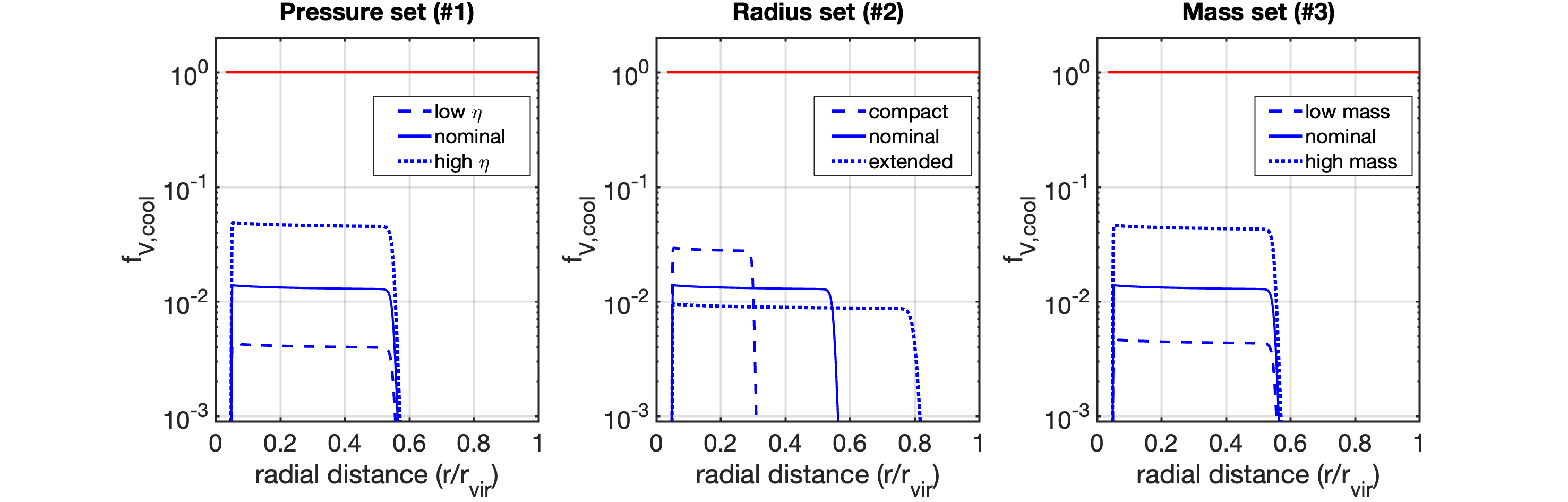}
 \caption{Cool gas volume filling fractions (unity is shown for scale by the red line). In the Pressure and Radius sets ($\#1$ and $\#2$, left and middle panels), the volume fraction is set to keep the cool gas mass constant between models (see Eq.~\ref{eq:mcool1}). In the Mass set ($\#3$, right panel) it is set to vary the cool gas mass by a factor of 3 ($\pm 0.5$~dex) from the nominal model (see Table~\ref{tab:inputs}).}
  \label{fig:vol_frac}
\end{figure*}

To allow for an easy comparison between the model sets, we define a nominal model -- a single combination of parameters chosen to reproduce the binned data. We then present two variations on this model in the parameter that is being explored in each set. This results in a total of 7 different models, and their input parameters and main outputs are summarized in Table~\ref{tab:inputs}. We plot the gas thermal properties and the volume filling fractions in Figures~\ref{fig:thermal} and~\ref{fig:vol_frac}, respectively. The plots show the profiles for the warm/hot (red) and the cool (blue) phases. We show the full profiles, extending out to $\rcgm$, for demonstration, and the thick part of each curve highlights the extent of the nominal model, $r_2=0.55 \rvir$.

As shown in Table~\ref{tab:inputs} and Figure~\ref{fig:vol_frac},  for the models presented in this section the volume filling fraction is $\approx 0.5-5\%$ of the total CGM volume where cool gas is present (and up to $10\%$ in the models discussed in \S\ref{sec:advanced}). These values are similar to the volume filling fraction reported by \cite{Ocker21} for the MW CGM, $\sim 1\%$, constrained by the amount of scattering measured for fast radio bursts (FRBs), and significantly higher than the fraction adopted by \cite{VP19}, $\fvc \sim 0.01 \%$. In \S\ref{subsec:disc_sizes} we discuss how this result can be used to constrain the sizes of cool gas clouds.

The full cool gas density profiles (middle panel of Figure~\ref{fig:thermal}) can be approximated as power-law functions of the distance from the halo center, $\ncool = \ncz (r/\rvir)^{-\anc}$, with $\anc>0$, and the fitted parameters are listed in Table~\ref{tab:inputs}. For the nominal model, $\anc = 1.67$, and the full range of slopes for the models presented here is $1.50 \lesssim \anc \lesssim 1.80$. These approximations are accurate to within $20\%$ of the numerical profiles for models with $r_2=0.55 \rvir$, and to $10\%$ and $40\%$ for the compact and extended models, respectively.

We now present and discuss each set of models in detail, examining the behavior of gas properties as functions of the model parameters, and comparing the resulting column density profiles, shown in Figures~\ref{fig:hi} (\HI), \ref{fig:carbon} (\CII-\CIV), and \ref{fig:silicon} (\SiII-\SiIV), to observations. We address the observed and model columns of \MgII~separately in \S\ref{subsec:magnesium} (and Figure~\ref{fig:mg2col}) since the constraints they provide on the gas properties are similar to those of \CII~and \SiII.

 \bgroup
 \def\arraystretch{1.2}
 \begin{table*}
 \centering
 	\caption{Model Properties}
 	\label{tab:inputs}
\begin{tabular}{| l || c | c | c | c | c |}
\toprule
\multicolumn{6}{| c |}{Input Parameters}	\\
\midrule \midrule
& Nominal & Pressure set ($\#1$, \S\ref{subsec:mod_pressure}) & Radius set ($\#2$, \S\ref{subsec:mod_radius}) & Mass set ($\#3$, \S\ref{subsec:mod_mass})	&  High Mass (\S\ref{subsec:adv_mod1}) \\
\midrule
  $\eta = \ac/\ah$    & \cc $3$    & \cc $1$, $9$      & \cc $3$	            & \cc $3$      & \cc $1$ , $5$      \tabularnewline
  $r_2/\rvir$         & \cc $0.55$ & \cc $0.55$        & \cc $0.30$ , $0.80$  & \cc $0.55$   & \cc $0.55$   \tabularnewline
  $\fvc/100$          & \cc $1.30$ & \cc $0.40$, $4.6$ & \cc $2.8$ , $0.90$   & \cc $0.45$ , $4.4$ & \cc $1.3$ , $7.8$ \tabularnewline
\midrule
\multicolumn{6}{| c |}{Key Output Properties} \\
\midrule
 \mcool/$10^{9}$~\msun& \cc $3$   & 	\cc $3$             & \cc $3$            & \cc $1$ , $10$ & \cc $10$ , $10$      \tabularnewline
 $\anc$               & \cc $1.67$& 	\cc $1.60$ , $1.74$ & \cc $1.52$ , $1.76$& \cc $1.67$     & \cc $1.60$ , $1.71$  \tabularnewline
 $\ncz/10^{-4}$       & \cc $2.3$ & 	\cc $8.1$ , $0.60$  & \cc $3.2$ , $1.9$  & \cc $2.3$      & \cc $8.4$ , $1.2$    \tabularnewline
\bottomrule
\end{tabular}
\end{table*}


\begin{figure*}[t]
\includegraphics[width=0.99\textwidth]{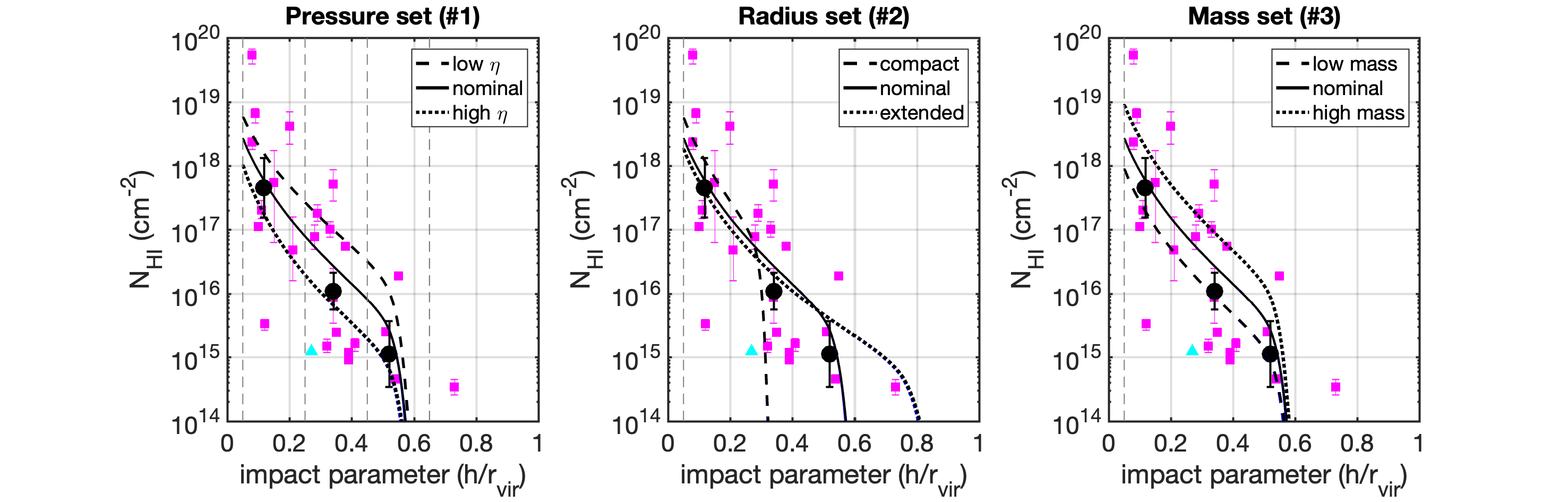}
 \caption{\HI~column densities for each of the scenarios described in \S\ref{sec:models}. {\bf Left:} In the Pressure set ($\#1$), the amount of non-thermal support affects the \HI~column density through photoionization, with lower densities leading to lower \HI~fractions. {\bf Middle:} Variation in the gas radial distribution (Radius set, $\#2$) affects the shape of the column density profile at large impact parameters. {\bf Right:} Variation in the gas mass through the volume filling fraction (Mass set, $\#3$) changes the column density without affecting the gas density and ionization state.}
 \label{fig:hi}
\end{figure*}

 \begin{figure*}[t]
\includegraphics[width=0.99\textwidth]{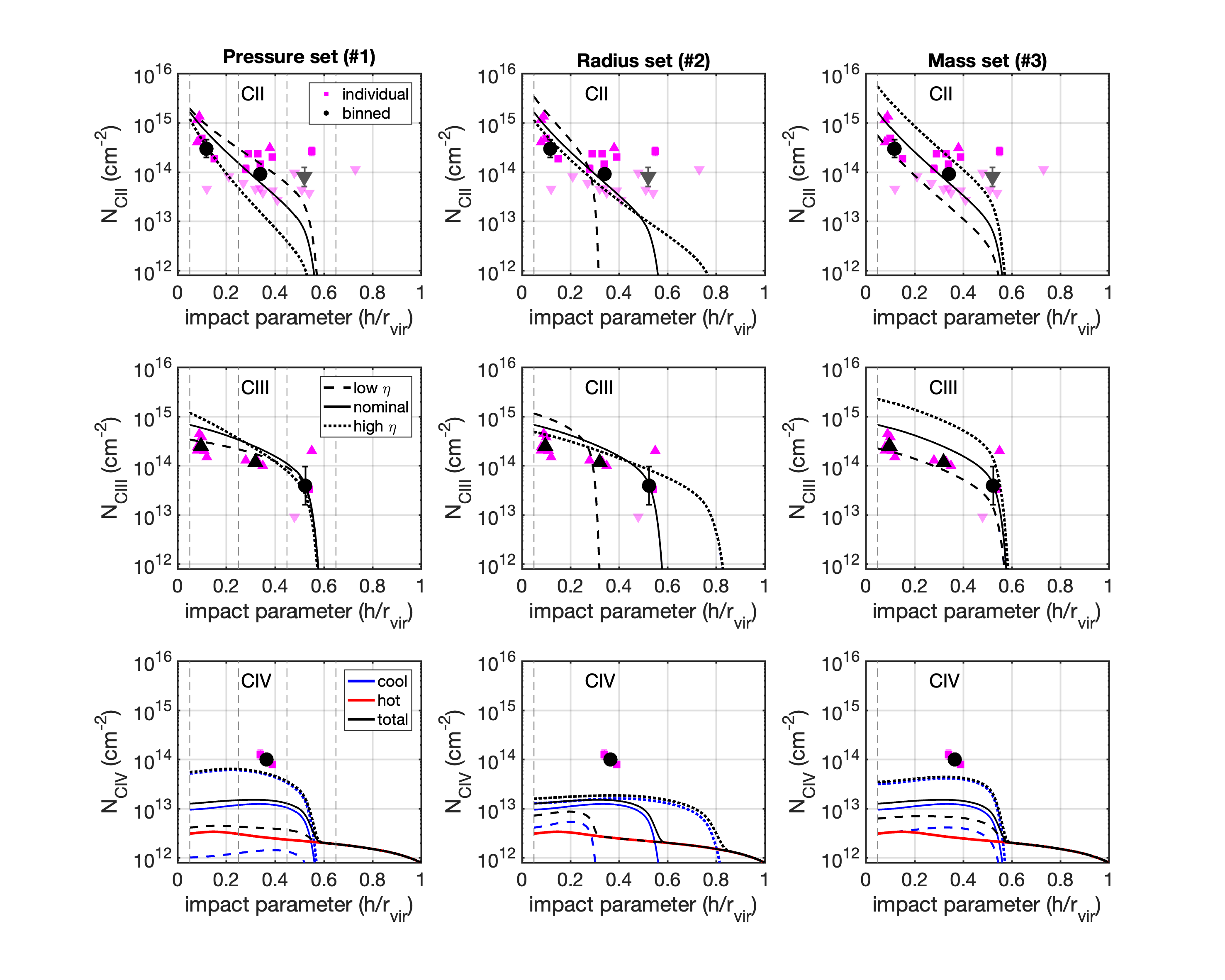}
 \caption{Carbon column densities - observations, individual and binned (magenta and black, respectively) and model results for the Pressure, Radius and Mass sets (left to right), for different ions - \CII, \CIII, and \CIV~(top to bottom, see \S\ref{sec:models} for details). Individual measured columns are shown by square markers, lower and upper limits are shown by up-pointing and faded down-pointing triangles, respectively. The vertical dashed lines show the bin boundaries and the innermost radius of the model, $r_1 = 0.05$\rvir. For some models, \CIV~(bottom panels) has a non-negligible contribution from the warm/hot CGM (red curve).}
 \label{fig:carbon}
\end{figure*}

\begin{figure*}[t]
\includegraphics[width=0.99\textwidth]{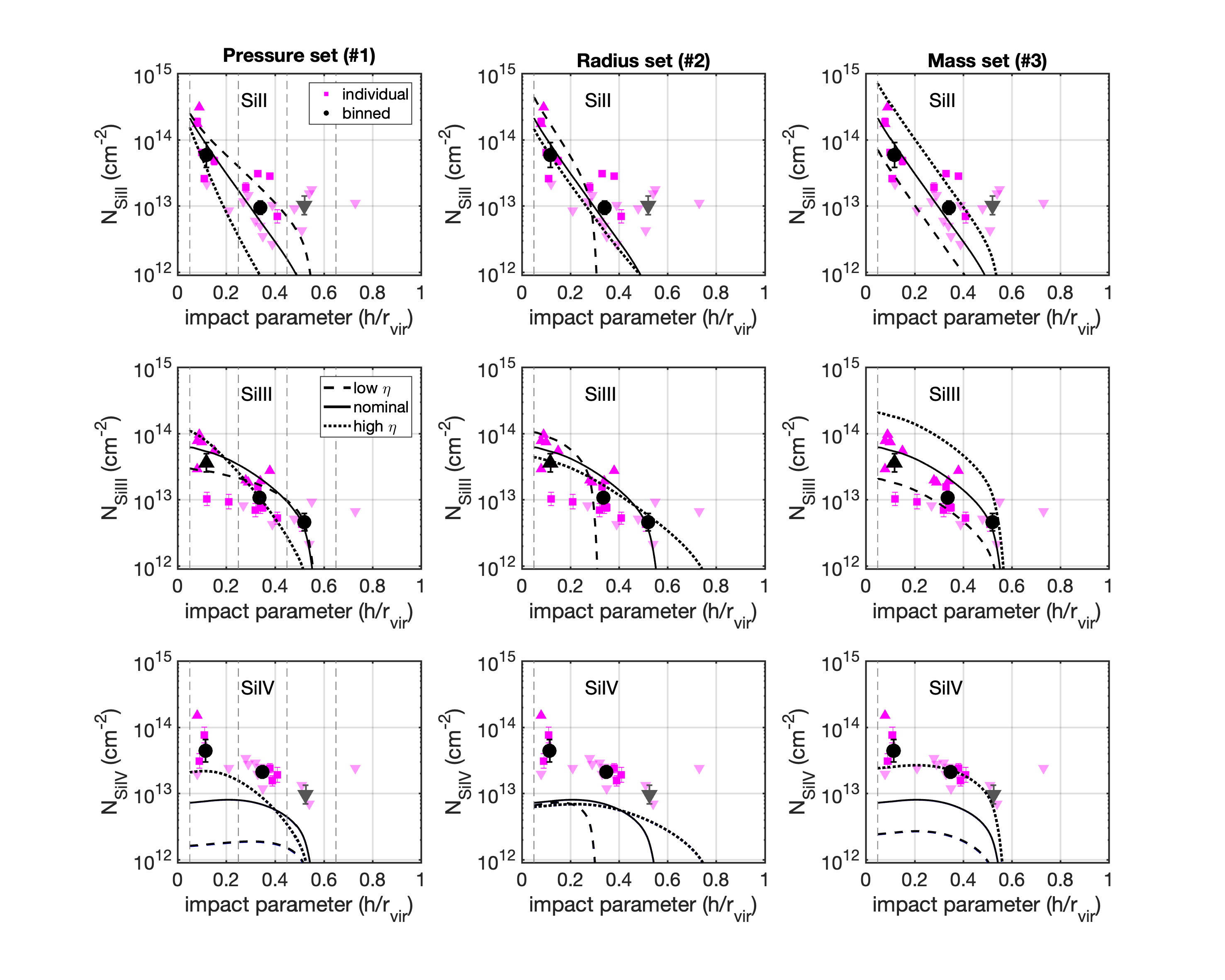}
 \caption{Silicon column densities - same as Figure~\ref{fig:carbon}, for \SiII, \SiIII, and \SiIV~(see \S\ref{sec:models} for details).}
 \label{fig:silicon}
\end{figure*}
 
\subsection{Set~$\#1$ - Non-thermal Pressure Support}
\label{subsec:mod_pressure}

In this set we vary the amount of non-thermal support in the cool gas and examine three values of $\eta$ -- $1$, $3$, and $9$, which we address as low, nominal, and high non-thermal support. The thermal properties for the three models are shown by the dashed, solid, and dotted blue curves, respectively, in Figure~\ref{fig:thermal}. The thick part of each curve denotes the range in which the cool gas volume filling fraction is non-zero, out to $r_2=0.55$~\rvir. The left panel of Figure~\ref{fig:thermal} shows $\ac = P_{\rm cool,tot}/P_{\rm cool,th}$ for these models, which varies with radius and is a factor of $\approx 1.5-3$ higher than $\eta$.

As shown by Equation~\eqref{eq:ncool}, for a given total pressure profile of the warm/hot gas, higher $\eta$ leads to lower gas volume densities (middle panel of Figure~\ref{fig:thermal}). The volume filling fraction in each model is adjusted to give $\mcool = 3.0 \times 10^{9}$~\msun, leading to higher \fvc~for models with higher $\eta$, and these are plotted in the left panel of Figure~\ref{fig:vol_frac}.

The temperature of the cool phase is set by heating/cooling equilibrium with the MGRF, and it is a function of the gas density and metallicity. For a constant metallicity, lower densities have lower net-cooling rates, leading to higher temperatures. This results in a temperature increase with radius, from $\tc \approx 10^4$~K at small radii ($<0.2$~\rvir), to $\approx 2 \times 10^4$~K at~\rvir~(right panel of Figure~\ref{fig:thermal})\footnote{~This is different from the warm/hot phase, for which \citetalias{FSM20} adopted the polytropic equation of state, resulting in higher temperatures for denser gas, and a temperature profile that decreases with radius.}. For the nominal model, extending out to $r_2 = 0.55$~\rvir, the temperature range is smaller. The variation in gas temperature between models is also small, with a maximal difference of $\lesssim 25\%$ at \rvir, compared to the nominal model, and it has a small effect on the ion fractions and column densities.

The gas ionization state changes with gas density and affects the ion fractions, total columns, and their ratios. We provide plots of the ion fractions and volume densities as functions of radius in Appendix~B (see Figure~\ref{fig:zap2_all}), and show how the ratio of ion column densities can constrain the amount non-thermal pressure support in \S\ref{subsec:adv_estim}. We now describe the behavior of the integrated column densities as functions of $\eta$, shown in the left panels of Figure~\ref{fig:hi} (\HI), Figure~\ref{fig:carbon} (\CII-\CIV, top to bottom), and Figure~\ref{fig:silicon} (\SiII-\SiIV).

First, for gas at $T \sim 10^4$~K and low densities ($\nh<0.1$~\cmv), \HI~is always removed by radiation. As a result, $N_{\HI}$ (left panel in Figure~\ref{fig:hi}) is higher in the model with low non-thermal support ($\eta=1$, dashed curve), as a result of higher gas densities and lower ionization state. Increasing $\eta$ leads to lower gas densities and \HI~fractions and columns (solid and dotted curves). \CII~and \SiII~(top left panels in Figures~\ref{fig:carbon} and \ref{fig:silicon}) behave similarly, with higher (lower) columns for lower (higher) values of $\eta$. Due to the difference in ionization potential between \CII~and~\SiII~($24.4$ and $16.3$~eV, respectively), photoionization has a stronger effect on \SiII\footnote{The flux at lower photon energies is higher, and the relative change in ionization parameter for a given change in density is larger.}.

For \CIII~and \SiIII~(middle left), the effect of photoionization changes at a threshold density of $n_{\rm H,thresh} \sim 5 \times 10^{-4}$~\cmv, from forming the ion at higher densities to removing it at $n_{\rm H} < n_{\rm H,thresh}$. In our models, the densities at $r< 0.4$~\rvir~are above this threshold, and lower gas densities in that region lead to higher ion fractions. As a result, the \CIII~and \SiIII~columns at small impact parameters are higher in the high-$\eta$ model (dotted curves). At larger impact parameters, photoionization has a small effect, and the column densities are similar for the different models.

Finally, \CIV~and \SiIV~(bottom left panels) are created by photoionization, and their columns are higher for the high-$\eta$ model, opposite of the \CII~and \SiII. The \CIV~and \SiIV~ion fractions are low at small distances from the galaxy (where the gas densities are higher) and high at larger distances, leading to column density profiles that are relatively flat out to $r_2$.

For most of the ions presented here, their column in the warm/hot phase is negligible compared to that in the cool phase ($<10\%$ for \SiIV, and below $1\%$ for all other ions) and does not appear in our plots. This is not the case for \CIV~(bottom left panel in Figure~\ref{fig:carbon}), and we plot the contribution from the warm/hot and cool gas with red and blue curves, respectively. In the warm/hot phase, \CIV~forms mainly at large radii, where the temperature is lower than in the central part of the CGM (see Figure~\ref{fig:thermal}) and closer to the optimal value for the ion, with a peak (CIE) fraction at $T \sim 10^5$~K\footnote{Since the gas temperature in the warm/hot phase is above the \CIV~peak temperature, and the ionization parameter at large radii is high, photoionization reduces the \CIV~ion fraction in the warm/hot phase, instead of increasing it, as it does in the cool phase.}. This leads to a flat column density profile, with $N_{\CIV}\sim 2\times 10^{12}$~\cmc. The column densities in the cool and hot phases are similar in the low-$\eta$ model, with $\sim 5 \times 10^{12}$~\cmc, and in the high-$\eta$ model, the cool phase dominates, with $N_{\CIV} \approx 4-5 \times 10^{13}$~\cmc.

We now compare our models to observations. For \HI, \CII, and \SiII, the nominal model is consistent with the binned data, and the range enclosed by the low and high-$\eta$~models encompasses $50-75\%$ of the individual measurements and limits. For the intermediate ions, \CIII~and \SiIII, the three models produce similar column densities. The nominal model is consistent with the ~\CIII~binned data, and a factor $\approx 2-3$ lower than the mean $\SiIII$~column at $h \approx 0.4$~\rvir. The detected column densities of the high ions, \CIV~and \SiIV, are underproduced by all three models. However, for \SiIV, all three models are consistent with the observed upper limits, which constitute $2/3$ of the measurements. The high $\eta$ model is a factor of $2-3$ lower than the binned $\SiIV$ data and the individual detections at $h \lesssim 0.4$~\rvir, with column densities of $\sim 3-8 \times 10^{13}$~\cmc. This model also predicts a total $\CIV$~column of $\sim 7 \times 10^{13}$~\cmc, within a factor of $\lesssim 2$ of the measured values at $h \approx 0.4$~\rvir, with $N_{\CIV} \sim 10^{14}$~\cmc. We discuss possible solutions for this tension in \S\ref{sec:advanced}.

\subsection{Set~$\#2$ - Radial Distribution}
\label{subsec:mod_radius}

In this set, we vary the outer radial boundary of the cool gas spatial distribution, and examine models with $r_2 = 0.30$, $0.55$, and $0.80$~$\rvir$, which we address as compact, nominal, and extended, respectively. The gas thermal properties do not vary with $r_2$ and are plotted as functions of radius by the solid curves in Figure~\ref{fig:thermal}. The volume filling fraction in each model is adjusted to give a gas mass of $\mcool = 3 \times 10^{9}$~\msun~(identical to Set~$\#1$), and they are plotted in the middle panel of Figure~\ref{fig:vol_frac}. The solid curve in each panel shows the nominal model, and the dashed and dotted lines - the compact and extended distributions, respectively.

The middle panel of Figure~\ref{fig:hi} shows the \HI~column densities for these three models, and carbon and silicon ions are shown in the middle column panels of Figures~\ref{fig:carbon} and \ref{fig:silicon}, respectively. For a given density profile, the shape and normalization of the column density profiles are determined by two factors: (i) the mean gas density, varying with the gas radial extent, sampling different parts of the full density profile, and (ii) the volume filling fraction, which for a given total gas mass is lower for more extended distributions (see middle panel of Figure~\ref{fig:vol_frac}). For example, the low ions -- \CII, \SiII~(top middle panels of Figures~\ref{fig:carbon} and \ref{fig:silicon}), and \HI~ -- form mostly in the inner region of the halo (see also Figure~\ref{fig:zap2_all} in Appendix~B), resulting in steep profiles with impact parameter. In the compact distribution, more mass is concentrated at these radii, leading to column densities that are higher by a factor of $2-3$ at small impact parameters, compared to the nominal model. The intermediate ions -- \CIII~and \SiIII~(center panels) -- form at larger distances and have flatter profiles. The high ions -- \CIV~and \SiIV~(bottom middle panels) -- are formed mainly in low-density gas, residing at large distances from the halo center. This results in flat column density profiles out to $r_2$, and lower columns in the more compact distributions.

As described in \S\ref{subsec:mod_pressure}, the total \CIV~column density has a non-negligible contribution from the warm/hot gas, with $N_{\CIV,hot} \sim 1-2 \times 10^{12}$~\cmc~(shown by the red curve in the bottom middle panel of Figure~\ref{fig:carbon}). The total \CIV~column density (black curve) is divided equally between the cool and warm/hot phases in the compact model, and dominated by the cool phase in the extended model. For the nominal model, the \CIV~column density in the cool component at $h<0.5 \rvir$ is higher by a factor of $3-4$ than in the warm/hot.

Compared to observations, the nominal and extended models are similarly consistent with the mean columns of low and intermediate ions, up to a factor of $\sim 2$. At impact parameters where the two models differ significantly, $h \gtrsim 0.5$~\rvir, the \CII~and \SiII~observations provide mostly upper limits, and both models are allowed by the data. Future measurements at these large radii may allow to better constrain the cool gas distribution (see also \citealp{Wilde21,KT22} for $\cgm2$~data). The compact model overproduces (underproduces) the binned data of the low and intermediate ions at small (large) impact parameters, below (above) $h \approx 0.3$~\rvir, and one can argue that it can be ruled out for the average profile. However, the \HI~measurements show a cluster of data at $0.3 < h/\rvir < 0.6$, with $N_{\HI} \sim 10^{15}$~\cmc, which are enclosed by the compact and extended model curves. This suggests that these low \HI~columns may be indicative of the spatial extent of the cool gas distributions in individual objects.

The observed column densities of the high ions, \CIV~and \SiIV, are underproduced by all three models, similar to the result of the Pressure set. The \SiIV~detections at $h \sim 0.4$~\rvir, with column densities of $\sim 2 \times 10^{13}$~\cmc, are higher than the models by a factor of $\gtrsim 4-5$. The \CIV~measurements at similar impact parameters, with $N_{\CIV} \sim 10^{14}$~\cmc, are about an order of magnitude higher than the total model columns.

Finally, we consider a variation in the inner boundary of the cool gas distribution. For the models presented here, we use $r_1 = 0.05 \rvir$. Adopting a value of $r_1 \sim 0.2$~\rvir, for example, for the nominal or extended models, produces a constant column density at $h<r_1$. For $\mcool=3 \times 10^{9}$~\msun~this results in low columns, inconsistent with the high measurements and lower limits at these impact parameters, and increasing \mcool~overshoots the low ion columns at larger $h$. Furthermore, the lack of cool gas in the vicinity of the galaxy, while not impossible, may be an unusual scenario. For example, in the MW, many of the high velocity clouds (HVCs), detected in $21$~cm, reside relatively close to the Galactic disk, at $d<20~$~kpc~\citep{Putman12}.

\subsection{Set~$\#3$ - Gas Mass}
\label{subsec:mod_mass}

In this set of models we vary the total gas mass in the cool component, by changing its volume filling fraction, $\fvc$~(see Eq.~\ref{eq:mcool1}), and these are plotted in the right panel of Figure~\ref{fig:vol_frac}. The masses of these models are $1$, $3$, and $10 \times 10^9$~\msun, and they are shown by the dashed, solid, and dotted curves, respectively. These masses constitute $\sim 1.8\%$ ($8.3\%$), $5.5\%$ ($25\%$), and $18\%$ ($83\%$) of the warm/hot CGM mass in the FSM20 fiducial model inside \rvir~($r_2$, and see Figure~\ref{fig:adv_estim2}). The gas volume density profile (middle panel in Figure~\ref{fig:thermal}) does not change between models, and the ion column densities are linear with \fvc~(see Eq.~\ref{eq:hi_col}-\ref{eq:met_col}). We now show that the range of columns produced by these models is consistent with a large fraction of the observations.

The \HI~column density profiles are shown in the right panel of Figure~\ref{fig:hi}, and encompass $1/2$ of the individual measurements. Data points outside the range predicted by the models are either low or very high columns. Sightlines with low  column densities, $N_{\HI} \sim 1-4 \times 10^{15}$~\cmc, may either belong to objects that are gas-deficient, or probe the outer boundaries of the cool gas distributions in the CGM, as suggested in \S\ref{subsec:mod_radius}. The high columns, $N_{\HI} \gtrsim 10^{18}$~\cmc, can be explained by models with higher gas density and higher mass cool phase (see \S\ref{subsec:adv_mod1}), or by large/high-density structures that intersect some lines of sight (similar to the Magellanic Stream in the MW, for example.).

The metal ion columns are shown in the right column of Figures~\ref{fig:carbon} and \ref{fig:silicon}, and the range predicted by this set of models is consistent with $> 3/4$ of the individual data points for the low and intermediate ions. The top right panels show the \CII~and \SiII~column densities, and the models are consistent with $20/25$ data points for each ion. The intermediate ions (middle right) in the model are also consistent with most of the measurements -- the \CIII~columns are at or above the observed lower limits, constituting $10/12$ of the data points. The \SiIII~models are consistent with $19/24$ of the data points. For \CIV~and \SiIV~(bottom right), the high mass model columns are below the binned data, by a factor of $\lesssim 3$.

\subsection{\MgII}
\label{subsec:magnesium}

\begin{figure*}[ht]
\includegraphics[width=0.99\textwidth]{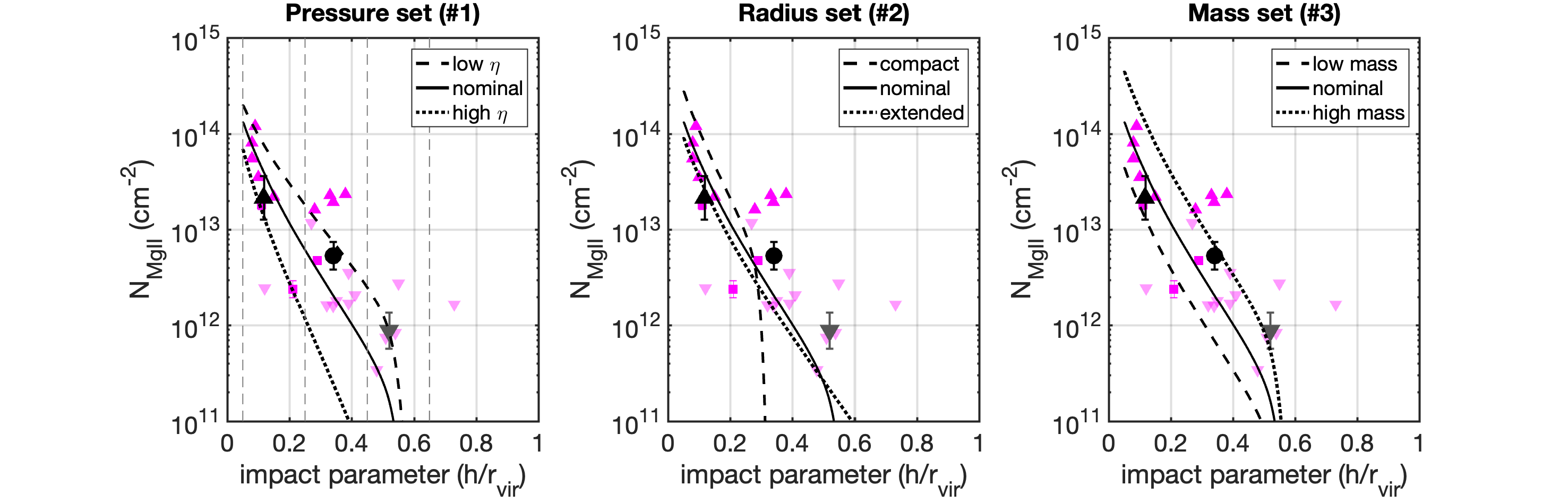}
 \caption{\MgII~column densities - observations and model results (see \S\ref{sec:models} and \S\ref{subsec:magnesium} for details.)}
  \label{fig:mg2col}
\end{figure*}

\MgII~is an ion that is often measured in the CGM \citep{Steidel92, Churchill00, Rigby02}. The doublet wavelength, at $\lambda=2796,2803A$, is close to the optical and can be probed from the ground at intermediate to high redshifts \citep{Kacprzak10, Matejek12, Evans13}. Figure~\ref{fig:mg2col} shows the \MgII~column densities measured for the COS-Halos sightlines with Keck/HIRES \citep{Werk13} and the column densities for the models described in \S\ref{subsec:mod_pressure}-\ref{subsec:mod_mass}.

The \MgII~ionization potential is $15.0$~eV, close to that of \SiII~($16.3$~eV), leading to similar column density profile shapes for the two ions in our models, with the \MgII~profiles slightly steeper with impact parameter. The solar abundance of magnesium is similar to that of silicon ($A_{Mg}/A_{Si} \approx 1.25$), and the absolute columns for the two ions are also similar.

The overall scatter in the observations is similar to that seen in \CII~and \SiII, and maybe slightly more bi-modal, with most of the reported measurements being upper or lower limits. Similar to the results presented in \S\ref{subsec:mod_radius}, the compact model underpredicts the column in the second radial bin. The models in the Pressure and Mass sets ($\#1$ and $\#3$) are consistent with a large fraction of the individual measurements or limits -- $17/25$ for $\eta$ variation, and $19/25$ for variation in \mcool.

\vspace{0.3cm}

To summarize this section, we presented a nominal model that is consistent with the binned measurements of low and intermediate ion column densities. We also explored how variation of each model parameter affects the cool gas properties and resulting column densities. We showed that models with $1<\eta<9$ ($2\lesssim \ac \lesssim 20$) produce column densities that are consistent with the binned measurements of the low and intermediate ions, and with $\approx 2/3$ of the individual detections and limits for these ions. Models with cool gas masses between $10^{9}$ and $10^{10}$~\msun~encompass $\approx 3/4$ of the individual measurements for the low and intermediate ions. Models with variation in the spatial extent of the distribution suggest that the cool gas of a typical galaxy extends to $\approx 0.6$~\rvir. However, low \HI~columns in individual objects may be indicative of outer boundaries between $0.3$ and $0.8$~\rvir. The model parameter combinations presented here underpredict the column densities of the high metal ions, \CIV~and \SiIV, by factors of $2-10$. Furthermore, about $20\%$ of the observed objects in our sample have \HI~columns higher than predicted by the models. In the next section we present models with variation in more than one parameter from the nominal model, and show these can reproduce the measured high and low column densities.

\section{Model Demonstration - Advanced}
\label{sec:advanced}

We now explore models that can produce columns lying beyond the typical scatter given by the models in \S\ref{sec:models}. First, in \S\ref{subsec:adv_mod1}, we explore cool gas models that combine variation in two parameters from the nominal values. We show that a high mass and low~$\eta$ combination gives high \HI~columns, and high mass and high~$\eta$ model reproduces the measured columns of high ions, \CIV~and \SiIV. Then, in \S\ref{subsec:adv_mod2} we address a scenario where the latter are produced in warm, intermediate temperature gas. We present the main outputs of these scenarios in Figure~\ref{fig:adv_mod}, and list their parameters and outputs in Table~\ref{tab:inputs}.

\begin{figure*}\label{fig:adv_mod}
\includegraphics[width=0.99\textwidth]{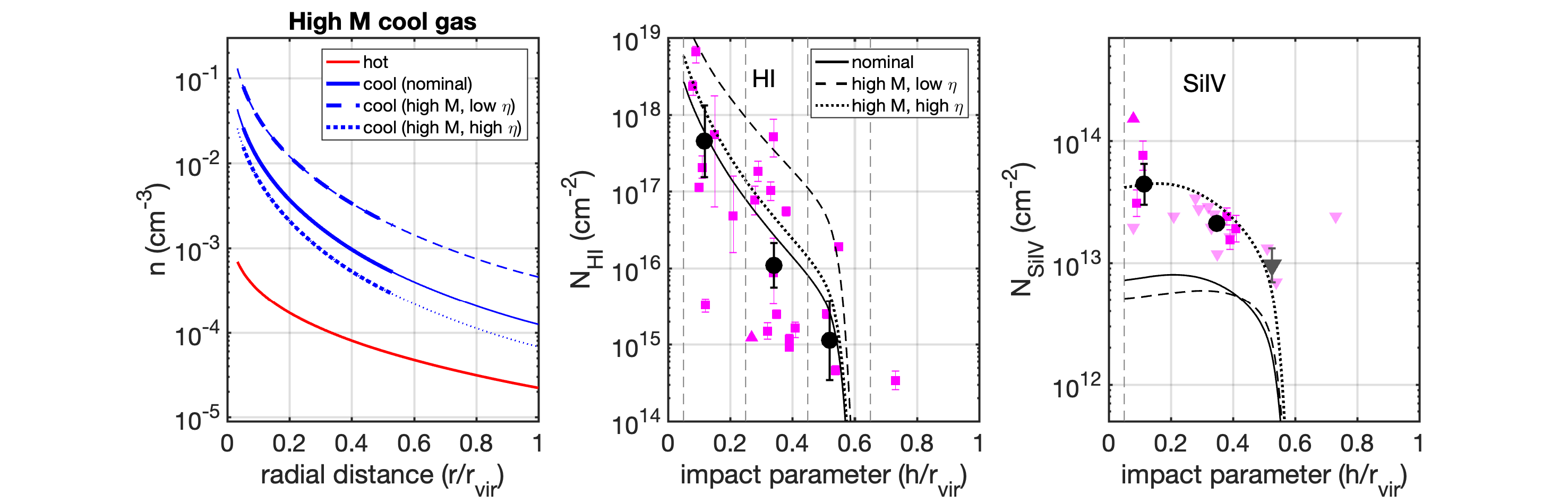} \\
\includegraphics[width=0.99\textwidth]{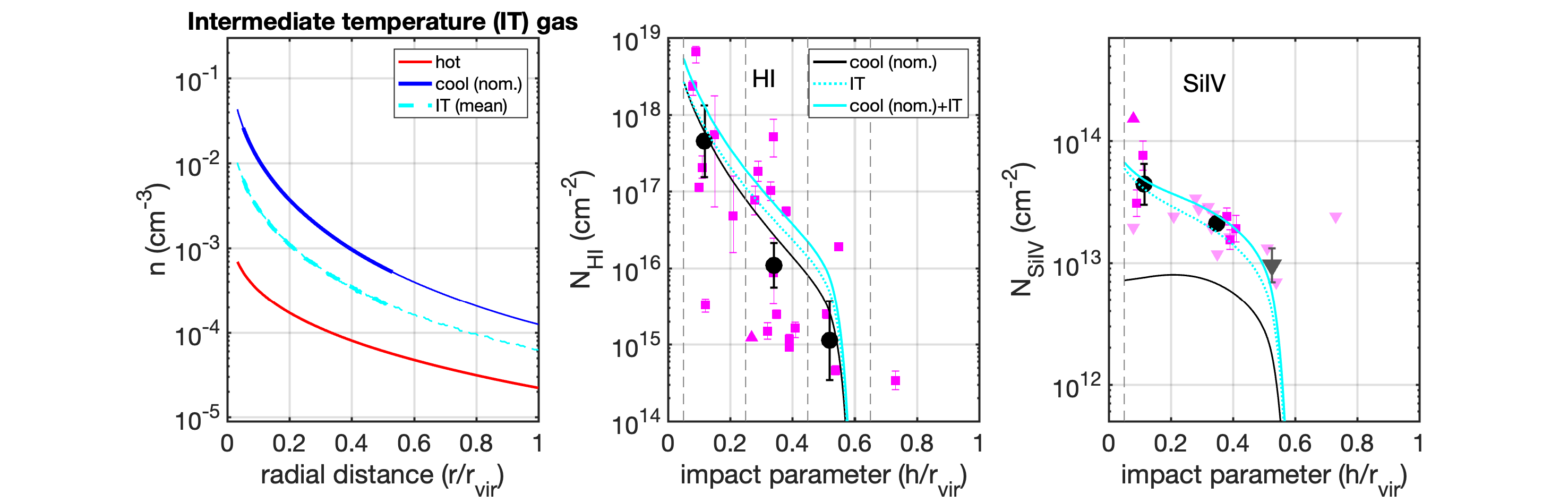}
  \caption{Advanced models (see \S\ref{sec:advanced}). {\bf Top:} Cool gas models with $\mcool = 10^{10}$~\msun. $\eta=1$ (dashed curves) gives higher volume densities (left), resulting in high \HI~columns (middle panel) and \SiIV~columns similar to the nominal model (right). $\eta=5$ (dotted) produces highly ionized gas, with high \SiIV, similar to the measured values, and nominal \HI~columns. {\bf Bottom:} Intermediate temperature gas (cyan dotted curves) with $M = 1.5 \times 10^{10}$~\msun~and $f_{\rm V} \approx 40\%$ can reproduce the measured \SiIV~column densities. Since this is a separate phase, the columns are added to those in the cool phase, and the total columns are shown by the cyan solid curves.}
\end{figure*}

\subsection{High Cool Gas Mass}
\label{subsec:adv_mod1}

In \S\ref{sec:models} we showed how variations in the model main parameters - non-thermal support, radial extent, and gas mass - affect the ion column density profiles. We addressed each parameter separately to better understand its effect, both qualitatively and quantitatively. For example, the effect of non-thermal support on the ion columns varies between the low and high ions - increasing $\eta$ reduces the local gas density, increases the gas ionization parameter, and leads to a higher ratio of the high to low ion column densities (see \S\ref{subsec:mod_pressure} and left panels of Figure~\ref{fig:carbon}). Varying the cool gas mass through the gas volume filling fraction, on the other hand, affects all columns in the same way, since $N_{\rm ion} \propto \fvc$ (\S\ref{subsec:mod_mass}, right panels).

We now show that models with higher mass, $\mcool = 10^{10}$~\msun, can reproduce either the high \HI~or \SiIV~(and \CIV)~observed columns, depending on the gas ionization state, set by the value of $\eta$. We examine two specific models, one with $\eta=1$, and the other with $\eta=5$, both with the nominal spatial extent, $r_2 = 0.55$~\rvir, and metallicity, $Z'=0.3$. We show some results of the two models in the top panels of Figure~\ref{fig:adv_mod} - the gas densities (left), the \HI~columns (middle), and the \SiIV~columns (right). We also plot the outputs of the nominal model from \S\ref{sec:models}, for comparison. The main inputs and outputs of the two models are also listed in Table~\ref{tab:inputs}.

The model with $\eta=1$ (dashed curves) has high volume gas densities (top left panel), with a profile identical to the low-$\eta$ model in \S\ref{subsec:mod_pressure}. This increase is enough to give the higher gas mass, and the volume filling fraction is similar to the nominal model, with $\fvc=1.3\%$. The higher density gas is less ionized, compared to nominal, and has higher \HI~fractions. The combination of higher gas densities and higher \HI~fractions leads to an increase by a factor of $\sim 10$ in the \HI~column densities (top middle), and these are comparable to the highest measured columns. The \SiIV~ion fractions, on the other hand, are lower than in the nominal model, canceling out the increase in $\nh$, and the resulting columns are similar to the nominal (top right).

The model with $\eta=5$ (dotted curves) has lower gas densities, and to produce the same gas mass requires a higher volume filling fraction, $\fvc=7.8\%$. The gas is more ionized, resulting in lower $\HI$ fractions, but the increase in $\fvc$ offsets the decrease in gas density and ion fraction and the $\HI$ columns are similar to those in the nominal model. The \SiIV~fractions are higher, and the resulting column densities are consistent with the binned data, and the individual measurements at $\approx 0.2$ and $\approx 0.4$~\rvir. This model also reproduces the observed \CIV~columns at $\approx 0.4$~\rvir~(not plotted here, see attached data files).

These results provide a general prediction from our model for the behavior of low and high ions in individual objects. For example, sightlines with high \HI~columns should have nominal columns of high ions (similar to nominal), and vice versa - sightlines with high \SiIV~columns should have nominal \HI~columns. The modeling of individual objects is beyond the scope of this paper, to be pursued in a follow up study.

Finally, a combination of low $\mcool$~with low~$\eta$ will produce typical \HI~columns and very low columns for high ions, and low $\mcool$~with high~$\eta$ will give very low \HI~and typical columns for high ions. These may also be interesting for individual objects or future observations with strong upper limits.

\subsection{Intermediate Temperature Gas}
\label{subsec:adv_mod2}

We now consider a second option, in which the high ions columns are formed in gas at intermediate temperatures, between that of the warm/hot and the cool phases. Possible scenarios for such models include gas that is cooling from the hot phase \citep{Heckman02,Qu18b}, or mixing gas at the boundaries of cool clouds \citep{Ji19,Gronke20,Fielding20a,Tan21a}.

To test this scenario, we model an intermediate temperature (hereafter IT) phase with a flat probability distribution, occupying the range between the temperatures of the warm/hot and the cool phases, $\th$ and $\tc$, at a given radius. We adopt a distribution function given by $P(T) \propto T^{-1}$, flat in logarithmic temperature bins, and verify that adopting $P(T) = const.$, flat in linear bins, gives similar results. We assume that the gas in this phase is isobaric, and the gas mass is then given by
\beq \label{eq:dist_mass}
\mint = 4 \pi \int_{r_1}^{r_2}{\mbar(r) \fvint(r) dr \int_{\tc}^{\th}{P(T) \nint(T) dT}}~,
\eeq
and the column density for a given ion can be written as
\beq \label{eq:dist_col}
N_{ion}(h) = \int_{s}{\fvint(r) A_{X} Z'(r) \int_{\tc}^{\th}{P(T) \nint(T) f_{\rm ion}(T) dT} ds}~,
\eeq
where $\fvint$ is the volume filling fraction of IT gas, taken to be constant with radius. We adopt the nominal extent and metallicity of the cool gas for the IT phase, $r_2 = 0.55~\rvir$ and $Z'=0.3$\footnote{~In general, the metallicity of the intermediate temperature phase is related to the origin of this phase (cooling, mixing, etc). Here we assume for simplicity that the metallicity of the intermediate phase is constant and equal to that of the cool gas, and a more detailed treatment can be undertaken in a future study.}, and vary the gas mass of the latter through $\fvint$, to reproduce the observed \CIV~and \SiIV~column densities.

The dotted curves in the bottom panels of Figure~\ref{fig:adv_mod} show a model with $\mint \approx 1.5 \times 10^{10}$~\msun. The mass-weighted mean density (dotted cyan curve, bottom left) is a factor of $\sim 2$ lower than the cool gas density in the nominal model (solid blue), and similar to that of the cool gas model with $\eta=5$ presented in \S\ref{subsec:adv_mod1} (dotted curves, top panels)\footnote{~This is not very surprising, since the gas density sets its ionization state, which is somewhat similar in the two models reproducing the measured columns of high ions. However, the two models do produce different column densities of low ions.}. The column densities for \HI~ (bottom middle) are also similar to the columns in the nominal model for the cool gas, and the total column (solid curve) are a factor of $\approx 2$ higher. The total \SiIV~column densities (bottom right panels) and the \CIV~(not shown) are dominated by the IT component.

To match the observed high ion columns, the model requires the IT phase to occupy a volume filling fraction of $\sim 40\%$, significantly larger than that of the cool gas, both in the nominal and high mass models, with $\fvc \approx 1\%$ and $\approx 10\%$, respectively. The resulting picture is different from simulation results (\citealp{Ji19,Ji20, Gronke20, Fielding20a,Tan21a,Tan21b,Gronke22}), who find thin mixing layers, occupying a small fraction of the total volume. This is also similar to the result by \cite{GS04}, who find that many boundary layers are needed to reproduce the observed \OVI~columns. The high $\fvint$ suggests our result may be more consistent with IT gas that cools from the warm/hot phase, possibly occupying a larger fraction of the total CGM volume.

We calculate how much \OVI~forms in the IT component, and find that the \OVI~column density profile has a similar shape to that of the $\SiIV$, decreasing from $\approx 10^{14}$~\cmc~in the inner part of the CGM, at $h/\rvir = 0.05$, to $\approx 10^{13}$~\cmc~at $h/\rvir \approx 0.5$. For comparison, the fiducial warm/hot gas model in \citetalias{FSM20} reproduces the observed \OVI~columns, $\approx 4 \times 10^{14}$~\cmc, approximately constant with impact parameter. Thus, while the \OVI~originating in the IT phase is not negligible at small impact parameters, it alone cannot reproduce the high measured columns at larger $h/\rvir$.

Finally, we note these results depend on the assumed metallicity for the IT phase. Increasing the metallicity by a factor of $2$ (to $Z'=0.6$) and keeping the same $\CIV$ columns will allow to reduce $\mint$, $\fvint$, and $N_{\HI}$~in this phase by a factor of $2$, resulting in a better agreement with the measured $\HI$~columns. However, the columns of low and intermediate metal ions and the volume filling fraction ($\approx 20\%$) will still be high compared to observations and mixing layers simulations, respectively. As noted above, the metallicity of the intermediate phase depends on the origin of the cool gas and on the small-scale physics in the boundary mixing layers. We leave a more detailed treatment of this issue to future studies.

\subsection{Estimating Gas Properties}
\label{subsec:adv_estim}

In Sections~\ref{sec:models} and \S\ref{subsec:adv_mod1} we presented models with specific parameter combinations and showed that by varying the non-thermal support and gas mass in the cool phase, our model framework allows to reproduce the mean observed column densities, the scatter in the data, and the high columns measured in some objects. We now perform a brief exploration of the model parameter space, and present outputs for a continuous variation of $\eta$ and $h$. We demonstrate how these can be used to easily relate the measured column densities to the properties of the cool CGM. As noted earlier, we leave the full modeling of individual lines of sight to a separate study.

\begin{figure*}[t]
\includegraphics[width=0.99\textwidth]{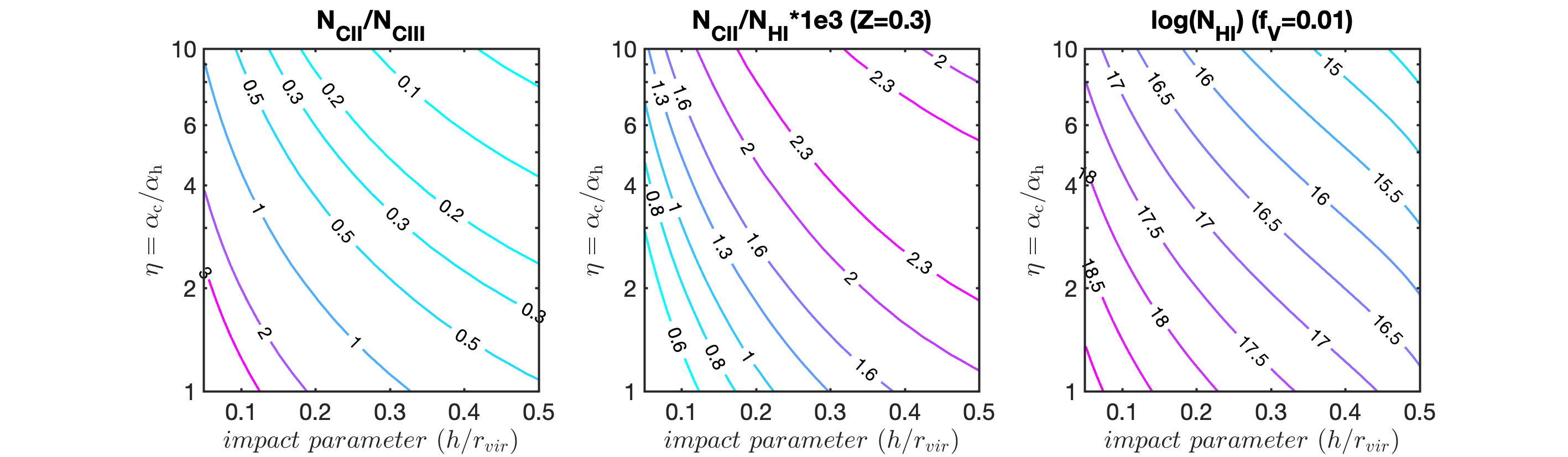}
 \caption{Estimating the cool gas parameters in individual lines of sight, using a model with $r_2 = 0.55$~\rvir~(see \S\ref{subsec:adv_estim} for details). {\bf Left:} The \CII/\CIII~column density ratio constrains $\eta$, the amount of non-thermal support. {\bf Middle:} The \CII/\HI~column density ratio allows to infer the gas metallicity. {\bf Right:} $N_{\HI}$ is a proxy for the cool gas volume filling fraction and mass (see also Figure~\ref{fig:adv_estim2}).}
  \label{fig:adv_estim1}
\end{figure*}

First, since in this work the gas metallicity $Z'$ and volume filling fraction $\fvc$ are constant as functions of radius, the metal column density ratios are independent of the assumed values for these parameters. For two ions of the same element, the ratio is also independent of the elemental abundance, and is only a function of the impact parameter and the amount of non-thermal support. As an example, in the left panel in Figure~\ref{fig:adv_estim1} we plot the $N_{\CII}/N_{\CIII}$ ratio in this 2D parameter space, for models with $r_2 = 0.55\rcgm$, and $\mcool= 3 \times 10^9$~\msun. The ratio of the measured columns for a given sightline allows to constrain the non-thermal support in the cool gas.

Second, the ratio of a given metal ion to \HI~columns scales linearly with metallicity, and independent of the volume filling fraction\footnote{If the model is used to reproduce only the metal column densities, without a measured hydrogen column, the gas metallicity and volume filling factor are degenerate, and only their product can be constrained.}. The middle panel plots the \CII/\HI~column density ratio in our model, for a metallicity of $Z'=0.3$, and for the \citealp{Asplund09} solar carbon abundance ($A_C = 2.7 \times 10^{-4}$). For an observed sightline, the metallicity is given by the ratio of the measured \CII/\HI~to the ratio in the plot, at the value of $\eta$ inferred in the previous step. The \CII~and \HI~fractions behave similarly with ionization parameter (i.e. gas density, see Figures~\ref{fig:hi}-\ref{fig:carbon}), and the overall variation in \CII/\HI~ratio with impact parameter and non-thermal support is relatively small. 

Finally, the total \HI~column scales linearly with the volume filling fraction and it is plotted in the right panel of Figure~\ref{fig:adv_estim1} for $\fvc = 1\%$. Similar to the metallicity, the volume filling fraction is given by the ratio of the measured \HI~column to the plotted value at a known $\eta$. At a given impact parameter, the \HI~column depends strongly on $\eta$, scaling as $\propto \eta^{-2.4}$. This strong dependence is a result of two factors - lower gas densities lead to smaller total hydrogen densities, and lower neutral fractions, due to the gas higher ionization state. 

\begin{figure}[ht]
\includegraphics[width=0.45\textwidth]{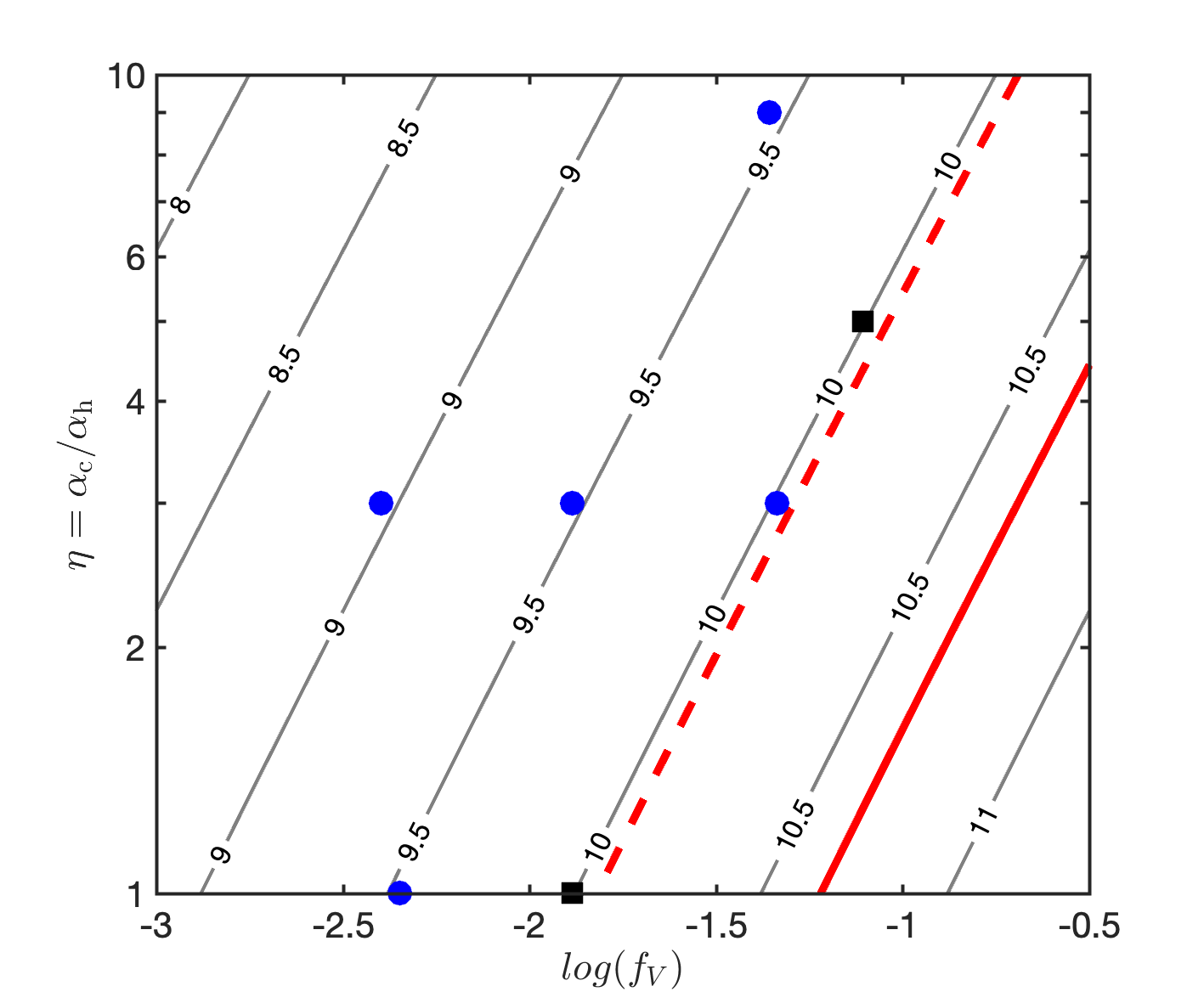}
 \caption{The cool gas mass as a function of the non-thermal support and volume filling fraction, for $r_2=0.55~\rvir$~(see \S\ref{subsec:adv_estim} for details). The markers show the individual parameter combinations adopted for the basic models discussed in \S\ref{sec:models} (blue circles), and the advanced models in \S\ref{subsec:adv_mod1} (black squares). The thick red contours indicates the warm/hot gas mass inside the spherical volume enclosed by $r_2$ (dashed) and inside \rvir~(solid), with $\mhot = 1.2 \times 10^{10}$ and $4.6 \times 10^{10}~\msun$, respectively.}
  \label{fig:adv_estim2}
\end{figure}

The non-thermal support in the cool gas and its volume filling fraction give \mcool. Inserting Eq.~\eqref{eq:ncool} into Eq.~\eqref{eq:mcool1}, the mass can be written as
\beq \label{eq:mcool2}
\mcool = \frac{\fvc}{\eta} \int_{r_1}^{r_2}{ \frac{\th}{\tc} dM_{\rm hot}} ~~~.
\eeq
In this study, we assume the fiducial FSM20 gas distribution for the warm/hot CGM, and $\th(r)$ is the same for all the models we examine (but see \S\ref{subsec:disc_sens} here). However, as shown in Section~\ref{sec:models}, the temperature of the cool component depends (weakly) on the gas density, which depends on the value of $\eta$. As a result, the gas mass deviates slightly from a linear function of $\eta$, and approximating it as a power-law we can write
\beq \label{eq:mcool3}
\mcool \approx 2.3 \times 10^{9} \times \left(\frac{\eta}{3}\right)^{-1.15}  \left(\frac{\fvc}{10^{-2}}\right) ~ \msun ~~~.
\eeq
We plot the total mass of cool gas, without approximation, in Figure~\ref{fig:adv_estim2} for $r_2 = 0.55$~\rvir. The individual markers show the models presented in \S\ref{sec:models} (blue circles) and in \S\ref{subsec:adv_mod1} (black squares). The thick red contours show the masses of warm/hot gas in the FSM20 fiducial model in the spherical volume enclosed by $r_2$ (dashed) and \rvir~(solid), with $\mhot = 1.2 \times 10^{10}$ and $4.6 \times 10^{10}$~\msun, respectively.

\section{Model Uncertainties}
\label{sec:uncertain}

We now address some of the assumptions of our model and the uncertainties of its components, and discuss their implications for the results presented in this work.

\subsection{Underlying Hot Gas Distribution}
\label{subsec:disc_sens}

The cool gas density in our model is set by the pressure profile of the warm/hot, volume-filling phase, and any profile for the latter can be used within our framework. In this work, we adopted the distribution from the \citetalias{FSM20}~fiducial model, and we now discuss how our main conclusions are affected by using a different pressure profile for the ambient medium. To address this, we look at the warm/hot gas density and pressure profiles from \citetalias{F22}, who considered two additional models with different amounts of non-thermal support (see their Section~3) - one with only thermal support in the warm/hot gas ($\ah=1$), and the second - dominated by non-thermal support ($\ah = 3$, and see their Fig.~6 for the gas thermal properties). We repeated our calculations for the cool gas nominal model parameters with these distributions, and we now describe the main findings of this analysis, also summarized in Table~\ref{tab:wh_sens}.

Before addressing the results, we note two points: (i) \citetalias{F22} show that models with low or high non-thermal pressure under- or over-produce, respectively, the measured \OVI~columns. This supports our choice of the \citetalias{FSM20} fiducial warm/hot gas parameters for this work, and the discussion here is mostly qualitative, to gain a better understanding of the cool gas model. (ii) The results presented here are valid at a given gas mass and boundary temperature in the warm/hot phase, and an exploration of the parameter space for a model with warm/hot and cool gas may be the focus of a future study.

To demonstrate the effect the warm/hot gas distribution has on the cool gas, we use Equation~\eqref{eq:ncool}: 
\beq
\ncool = \frac{P_{\rm th,hot}/\kb}{\eta \tc} = \frac{\ah}{\ac} \frac{P_{\rm th,hot}/\kb}{\tc} ~~~.
\eeq
We address how variation in $\ah$, the amount of non-thermal support in the warm/hot gas affects its pressure profile, and then how this affects the cool gas properties for two sub-cases - (a) fixing the value of $\eta$ to that of our nominal model, i.e. a constant ratio between the non-thermal pressures in the warm/hot and cool phases, and (b) fixing $\ac$ to the values in the nominal model, i.e. a given ratio of non-thermal to thermal pressure in the cool phase. These two cases may correspond to (a) cool gas forming by condensation from the warm/hot phase with non-thermal pressure from magnetic fields, for example, and (b) cool gas originating from `external' sources, such as outflows, dwarf galaxies, or IGM accretion.

First, for warm/hot gas with only thermal support ($\ah \approx 1$), at a given total gas mass and fixed thermal temperature at the outer boundary, the warm/hot gas temperature and density profiles are steeper than in the fiducial model, leading to a steeper thermal pressure profile, and higher pressures at radii $<0.6$~\rvir. For a fixed value of $\eta$, these higher pressures lead to higher cool gas densities. For a fixed value of $\ac$, the lower $\ah$ increases $\eta$, negates the effect of higher $P_{\rm th,hot}$, and leads to cool gas densities similar to the nominal model.

Second, for warm/hot gas with higher non-thermal support ($\ah \approx 3$), the temperature and density profiles are flatter than the fiducial FSM20 distributions, leading to lower pressures at $r<0.6$~\rvir. For a fixed $\eta$, lower $P_{\rm th,hot}$ lead to lower cool gas densities. For a fixed $\ac$, the result is lower $\eta$, and cool gas densities that are similar to the nominal model.

To summarize our results qualitatively, we find that for a given value of $\eta$, lower (higher) non-thermal support in the warm/hot phase, leads to higher (lower) ambient gas pressure at $r<0.6$~\rvir, and results in higher (lower) cool gas densities. The temperature of the cool gas is also slightly decreased (increased), enhancing the density change compared to nominal. On the other hand, for a fixed $\ac$ profile, lower (higher) non-thermal support in the hot gas gives higher (lower) $\eta$, opposing the effect of higher (lower) ambient gas pressures, and producing cool gas densities similar to those in the nominal model.

\bgroup
 \def\arraystretch{1.2}
 \begin{table}
 \centering
 	\caption{Variation in non-thermal support in the warm/hot phase}
 	\label{tab:wh_sens}
 		\begin{tabular}{| l || c | c |}
 			\midrule
 			& Thermal ($\ah \approx 1$) & Non-thermal ($\ah \approx 3$)  \\
 			\midrule
 			$P_{\rm th,hot}$ - profile shape    & steeper   & flatter     \\
 			$P_{\rm th,hot}(r<0.6~\rvir)$       & higher    & lower       \\
                (a) fixed $\eta$                    & higher $\ncool$   & lower $\ncool$	\\
 			(b) fixed $\ac$                     & similar $\ncool$  & similar $\ncool$  \\
 			\bottomrule
 	\end{tabular}
 \end{table}

\subsection{Radiation}
\label{subsec:disc_rad}

In this work we assume that the heating and photoionization of the cool CGM are dominated by the MGRF, and adopt the \citetalias{HM12} field. In this section we first discus our choice of the MGRF, and then examine a possible contribution from galactic radiation. For our discussion we address the spatially averaged ionizing photon flux ($E>13.6$~eV), given by $\Phi = \int{J_{\nu}d\nu}$ and depending on the radiation field intensity and spectral shape.

\subsubsection{Metagalactic Radiation Field (MGRF)}
\label{subsec:disc_rad_uvb}
For this work, we focus on the ionizing flux at $z<0.5$, dominated by photons at $13.6<E/{\rm eV}<100$. In this redshift and energy range, the MGRF is not well-constrained empirically, and existing works have different predictions for the field spectral shape and intensity. For example, the fields calculated by \citet{FG09} and \citet[hereafter KS19]{KS19} predict $\Phi$ values that are higher than HM12 by a factor of $\approx 1.5$ (see their Fig.~7 for a comparison). In this work we adopt the \citetalias{HM12} for consistency with \citetalias{FSM20} and verify that using the \citetalias{KS19} field changes the resulting ions fractions by $<50\%$.

How will a higher $\Phi$~affect our inferred model parameters? For a given hot gas pressure profile, reproducing a measured column density ratio with higher ionizing flux would require increasing the (mean) gas volume density by a similar factor (see also \citetalias{Prochaska17} for discussion). In our model, this can be achieved by reducing the non-thermal support, $\eta$, by the same amount (see Equation~\ref{eq:ncool}). Producing the same total ion columns with higher gas densities will require lowering the cool gas volume filling fraction, \fvc, and he total cool gas mass will be only weakly affected (see Equation \ref{eq:mcool3}).

In this work we also adopt a single ionizing flux, $\phi = 2 \times 10^{4}~{\rm ~photons~s^{-1}}$, corresponding to $z=0.2$, the median galaxy redshift in our sample. Models show that at low $z$, the MGRF flux is a strong function of the redshift, following the decline in the cosmic SFR and SMBH activity since $z \sim 2$. For the \citetalias{HM12} field at $z<0.5$, the flux evolution can be approximated by $\Phi \approx \Phi_0 \times (z+1)^{4}$, corresponding to an increase of $\approx 3$ from $z=0.1$ to $0.4$, the full redshift range in our sample. This variation in ionizing flux can be one of the factors contributing to the scatter between the observed columns, and will be addressed in modeling of the individual lines of sight.

\subsubsection{Local/Galactic Radiation}
\label{subsec:disc_rad_gal} 

In this work we assumed that the ionization of cool gas in the CGM is dominated by the MGRF. We now discuss a possible contribution of galactic radiation, from an AGN, stars, stellar remnants, or outflows, to the ionization of the CGM.}

Radiation from local sources is expected to fall off rapidly with the distance from the galaxy, as $r^{-2}$. This suggests that at small enough distances from the galaxy, galactic radiation will eventually become strong enough to significantly affect the gas ionization state (see \citealp{SMW02}). Where does this happen? \cite{MW18} estimated that local radiation sources (excluding AGN) cannot be dominant beyond $r \sim 100$~kpc from the galaxy. \cite{Upton18} add emission from quasars to the background field, and find that at $z \sim 0.2$ the ``proximity radius'', at which galactic and extragalactic background sources contribute equally to the local radiation field, is $\sim 10-30~{\rm kpc} \times \left( {\rm SFR/\msuny} \right)^{1/2}$. \cite{Holguin22} used hydrodynamic simulations and ionization modeling to show that stellar radiation has a negligible effect on the cool CGM at $r \gtrsim 30$~kpc ($\approx 0.1$~\rvir) in MW-mass halos at low redshift. Radiation from an accreting SMBH can be highly non-isotropic and significant enough to affect gas at large distances from the galaxy. The COS-Halos galaxies do not harbor an AGN, but \citealp{Opp18a} find that radiation can have a lingering effect on the \OVI, even after the AGN has switched off. Estimating the time-dependent effect of AGN radiation on gas ionization is beyond the scope of this work.

\cite{Sarkar22} perform simulations of winds driven by SNe, including radiative transfer and non-equilibrium ionization effects. They compare the emission from the winds to the intensity of the MGRF, and find that close to the star forming regions, $d<1$~kpc, wind radiation can be orders of magnitude stronger than the background flux (see their Fig.~3). However, at $r\sim 10$ kpc, the radiation intensity is close to the background, and extrapolating to $r \sim 30$~kpc, we expect the two components to be similar. Their simulations are run with $SFR=10$~\msuny, and they compare to the $z=0$ \citetalias{HM12} MGRF. Scaling these to the median SFR of the COS-Halos sample, $SFR=4.3$~\msuny, and the $z=0.2$ background flux should further lower the ratio of wind to background radiation by a factor of $\sim 5$, and we estimate that radiation from winds should be sub-dominant at $r \gtrsim 14$~kpc, or $\approx 0.05$~\rvir, and possibly even closer to the galaxy. 

We summarize that galactic radiation should not be dominant beyond $r \approx 0.1$~\rvir, and our approximation that the ionization is dominated by the MGRF at large radii is reasonable. This radius may be larger for individual galaxies, with higher present or recent SFR, or past SMBH activity, and these may contribute to the scatter in the data at small impact parameters.

\section{Discussion}
\label{sec:disc}

The model presented in this work, allows, when applied to observations, to constrain the mass, metallicity and spatial distribution of the cool CGM (\S\ref{sec:models} and \S\ref{subsec:adv_estim}). While the model itself is intentionally agnostic to the source of the cool gas or its formation mechanism, the results of our analysis may favor, or be more consistent, with some mechanisms over others. We now discuss constraints on the cloud sizes from our modeling results, the depletion time of cool CGM, and possible channels for its buildup or replenishment.

\subsection{Cool Gas - Cloud Sizes}
\label{subsec:disc_sizes}

Our model constrains the volume filling factor of the cool CGM, and does not address the sizes of individual clouds. We now show that combining the model volume filling factor with the sky covering fraction estimated from observations allows us to place some constraints on cloud sizes.

The volume number density of cool clouds at some distance from the galaxy can be written as by $\ncl = (dV_{\rm cool}/V_{\rm cl})/dV = \fvc/V_{\rm cl}$, where $V_{\rm cl}$ is the volume of a single cloud, $dV_{\rm cool}$ and $dV$ are the cool gas and total volume in a shell at some radius, and we used the definition of $\fvc$ (Eq.~\ref{eq:fvcool}). The number of clouds along a line of sight through the CGM is then
\beq \label{eq:Nls}
\nls(h) = \int_{S}{A_{\rm cl} \ncl ds} = \int_{S}{\frac{A_{\rm cl}}{V_{\rm cl}}\fvc ds} = \frac{3}{4} \int_{r_1}^{r_2}{\frac{\fvc}{\rcl}}ds ~,
\eeq
where $A_{\rm cl}$ is the cloud cross section, and for the last equality we assumed spherical clouds.

For simplicity, we assume a cloud size that is constant with radius, and in Figure~{\ref{fig:morphology} we plot $\nls$ as a function of the impact parameter, for cloud sizes of $\rcl = 0.2$, $0.6$, and $2$~kpc (dashed, solid, and dotted black curves, respectively), for our nominal model, with $\fvc \approx 1\%$. For $\rcl=0.6$~kpc, the number of clouds decreases weakly from $\nls\approx 5$ at $h=0.1$~\rvir~to $3$ at $h=0.4$~\rvir~and steeply after that. For $\rcl=0.6$~kpc we also plot $\nls$ for the extended model (solid blue curve).

\begin{figure}[h]
\includegraphics[width=0.45\textwidth]{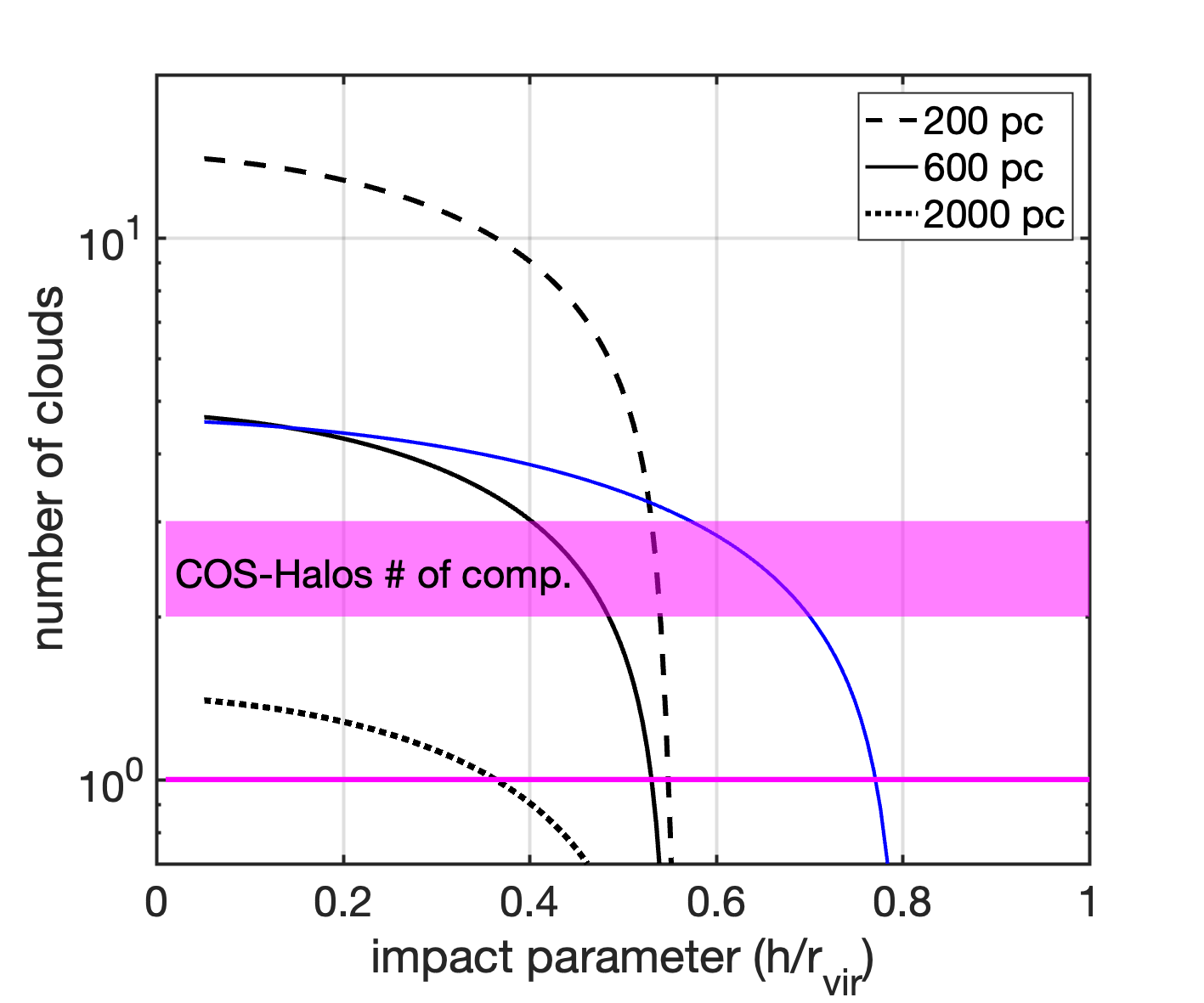}
 \caption{Number of clouds along a line of sight as a function of the impact parameter, for different clouds sizes. The black curves show the distribution for the nominal model, and the solid blue - for the extended model (see \S\ref{sec:models}). The horizontal magenta lines show $\nls=1$ (solid) as an approximate threshold for high detection rate, and $\nls \approx 2.4$ (shaded band) - the average number of spectral components per line of sight in COS-Halos \citepalias{Werk14}.}
  \label{fig:morphology}
\end{figure}

For a small number of clouds, we expect the detection rate to vary significantly between lines of sight, and for $\nls < 1$ (shown by the solid magenta line) we also expect the detection rate to be low. However, \citetalias{Werk14} report a detection rate, or sky covering fraction, close to unity for all impact parameters. Furthermore, they estimate an average of $\approx 2-3$ number of spectral components per line of sight in the COS-Halos data. This range, shown by the magenta band, can be taken as a lower limit on the number of clouds per line of sight, due to limited spectral resolution of COS and possible clustering of individual gas clouds in the CGM. For $\rcl=const.$, we can re-write Eq.~\eqref{eq:Nls} as a constraint on cloud size
\beq\label{eq:rcl_lim}
\begin{split}
R_{\rm cl}(h)
& = \frac{3 r_2\fvc}{2\nls}\sqrt{1-h^2/r_2^2} \\
& \lesssim 0.40 \left(\frac{\fvc}{0.01}\right) \left(\frac{\nls}{3}\right)^{-1} \left(\frac{r_2}{150~{\rm kpc}}\right)~{\rm kpc} ~~~,
\end{split}
\eeq
estimated at $h=0.85 r_2$ (for $\rvir=260$ kpc), and a factor of $\approx 2$ larger at $0.15$~\rvir. We conclude that for our nominal model, $\rcl \gtrsim 1.5$~kpc  predicts less than one cloud per line of sight and is inconsistent with the observed high detection rates, and for $N_{\rm cl} \gtrsim 3$, $\rcl \lesssim 0.5$~kpc.

For a given total mass in cool gas, variation in $\eta$ will affect the number of clouds through the volume filling factor. As discussed \S\ref{sec:models}, higher $\eta$ leads to lower gas density and requires higher \fvc~(see left panel of Figure~\ref{fig:vol_frac}, and Equation~\ref{eq:mcool2}). For the low (high)~$\eta$ model presented in this work, the number of clouds of a given size will be a factor of $\approx 3$ lower (higher), making the constraint on cloud size stronger for lower values of non-thermal support.

The mass of an individual cloud scales as the local gas density, and can be written as
\beq \label{eq:Mcl}
\Mcl(r) = 1.67 \times 10^4 \left(\frac{r}{100~{\rm kpc}}\right)^{-\anc} \left(\frac{\eta}{3}\right)^{-1} \left(\frac{\rcl}{0.5~{\rm kpc}}\right)^{3} \msun  ~~~,
\eeq
where $\anc\approx 1.67$ for our nominal model (see Table~\ref{tab:inputs}).

\citetalias{Werk14} estimate the clouds sizes as the ratio of the total gas column to the local density, $l = N/n$, and find a median of $10_{-10}^{+35}$ kpc, with a wide range, $0.1-2000$~kpc. Since this estimate assumes a single cloud for each line of sight, it provides an upper limit for the cloud size, and the median is consistent with our result. They then infer the gas volume filling fraction using the cloud size (assuming spherical geometry) and the sky covering fraction, adopting the method described by \cite{Stocke13}, and infer $\fvc \approx 11_{-9}^{+15}\%$. While these are higher than the values we infer, they are consistent within the estimated errors with our range, of $\approx 0.5-10\%$. For comparison, \cite{Stocke13} examine a different data set and estimate a median cloud size of $1$~kpc with a $\pm 1$~dex range.

The upper limit we estimate for $R_{\rm cl}$ can be compared to the results by \citet{Zahedy21}, who use absorption observations of lensed quasars to probe coherence of cloud properties across small spatial scales. They find that at separations below $\sim 0.15$~kpc, there is little variation in the column densities and velocities of absorbers (see also \citealp{Rubin15,Rubin18b}, \citealp{Rudie19}, and \citealp{Augustin21} for similar studies at higher redshifts). Our constraint is consistent with their result. If cloud sizes are indeed close to $0.15$~kpc, the number of individual clouds along a line of sight in our nominal model is $\approx 10$.

\cite{McCourt18} argued that gas undergoing thermal instability and cooling fragments (or `shatters') into smaller and smaller clouds, and the cloud size at the end of this process is given by the cooling length
\beq\label{eq:cscool}
\begin{split}
L_{\rm cs}
& = c_s \tcool = \left(\frac{\kb T}{\mbar} \right)^{1/2} \frac{n \kb T}{n_e n_H \Lambda} \\
& \approx 90 \left( \frac{n}{10^{-3}~\cmv} \right)^{-1} \left( \frac{T}{10^{4}~{\rm K}} \right)^{3/2} ~ {\rm pc} ~~~,
\end{split}
\eeq
where in the last approximation we used $\Lambda = 10^{-23}~{\rm erg~s^{-1}~cm^{3}}$, the net cooling efficiency of gas at $T \sim 10^4$~K. Equation~\eqref{eq:cscool} shows that for the densities of the cool phase in our models ($n \sim 10^{-4}-10^{-2}$~\cmv, see middle panel of Figure~\ref{fig:thermal}), clouds in the CGM have sizes of $\sim 10-1000$~pc, significantly larger than the subparsec-scale clouds estimated by this equation for ISM densities. Our limit from the observed gas covering fraction in Equation~\eqref{eq:rcl_lim} agrees with this size estimate at $h \lesssim 0.6$~\rvir.

The analysis presented here is highly simplified. The sizes of clouds do not have to be constant with distance from the galaxy if the properties of the ambient medium and the cool gas vary significantly. Cloud sizes may also be set or affected by additional physical processes other than cooling - turbulence and instabilities, for example, can disrupt clouds and break them to smaller sizes \citep{Arm17}, while magnetic fields may protect the clouds from fragmentation \citep{Sparre19}. Given these different mechanisms, the assumption of spherical geometry is also a simplification, and clouds that are infalling or have magnetic fields may be stretched into filaments. Finally, clouds may have a distribution of sizes, possibly related to their different origins (see \S\ref{subsec:disc_res}), and they may cluster and form complexes and larger structures, such as those observed in the MW \citep{Putman12} and suggested by higher resolution spectra \citep{Tripp22}. Given this potential complexity, we find the agreement between our results and other (observational and theoretical) works encouraging, and leave a more detailed study of cloud sizes, shapes, and kinematics for future study.

\subsection{The Cool CGM Reservoir - Depletion and Formation}
\label{subsec:disc_res}

The dense cool gas clouds are not supported by the (hydrostatic) pressure gradient of the ambient gas, and may fall towards the galaxy. For a cloud starting at rest at a distance $r$ from the halo center, the free-fall time is given by
\beq\label{tinf}
\tinf(r) \approx \left( \frac{\pi^2 r^3}{8G M(<r)} \right)^{1/2} \approx 2.3 \times 10^9 \left( \frac{r}{\rvir} \right)^{1.13} ~~ {\rm yr} ~~~,
\eeq
where we used the approximation for the \citetalias{Klypin02} profile, and $\rvir = 260$~kpc\footnote{\citetalias{FSM20} adopt the dynamical time used by \cite{Voit19}, $\tdyn = \sqrt{2r^3/G M(<r)}$, a factor of $\approx 1.3$ longer}. This is similar to the result obtained by \cite{Shull20} for a pure NFW halo, with $\tinf \sim 190$~Myr at $r=50$~kpc. This estimate is a lower limit on the infall time for clouds starting at rest, since it does not include the effect of ram pressure from the hot medium, which will slow the clouds down \citep{Afruni21}. Furthermore, clouds may be disrupted by hydrodynamical instabilities (see \citealp{Joung12b}, for example), or grow through cooling-driven accretion from the hot phase and slow down \citep{Tan22}.
Thus, the accretion process of cool gas from the CGM onto the galaxy is highly uncertain. For our analysis here, we can use \tinf~to estimate the mean accretion rate of cool gas, given by $\dmcool \sim \mcool/\tinf$. For the nominal model presented in \S\ref{sec:models}, with $\mcool=3 \times 10^9$~\msun~and $r_2=0.55$~\rvir, this gives $\dmcool \sim 3.1$~\msuny, and $\sim 1-10$~\msuny~for the models in the Mass set. The COS-Halos star-forming galaxies have star-formation rates in the range of $0.6-19$~\msuny, with a median of $4.3~\msuny$~(\citealp{Werk13}, and Table~\ref{tab:obscol} here), and we find this agreement reasonable. We note that our nominal result is also similar to the MW SFR, with $\approx 1.6-2.0$~\msuny~\citep{Chomiuk11,Lic15}. \cite{Fox19} estimate an inflow rate of $\sim 0.5$~\msuny~from observations of High Velocity Clouds (HVCs) in the MW, and comment that the actual mass inflow rate may be higher due to gas at lower velocities.

The high detection rate of cool CGM in the COS-Halos survey suggests that the reservoir of cool gas in the CGM is long lived, rather than a transient phenomenon. This can happen if accretion occurs on timescales significantly longer than the dynamical time (as noted earlier), or the cool gas is continuously replenished, resulting in a low net depletion rate. We now discuss the latter scenario.

Cool gas can form or be added to the CGM by different mechanisms, including accretion from the IGM \citep{Keres05,Dekel09,Wright21}, stripping from satellite galaxies \citep{Tonnesen09,Putman21}, condensation from the warm/hot ambient medium \citep{Sharma12a, Joung12a, Voit19}, and outflows from the galaxy \citep{Li20a,Fielding22}. We now address the possible contributions from these channels.

\cite{Correa15b} model cosmological accretion and find that for a $10^{12}$~\msun~halo at $z\sim 0.2$, baryons are added to halos at a rate of $\dot{M}\sim 8.2$~\msuny, linear with $\mvir$ (see their Eq.~23). \cite{Fakhouri10} report similar results, with slightly higher accretion rates (up to $\approx 40\%$). For massive halos at low $z$, we expect the infalling gas to shock around the virial radius and contribute mainly to the warm/hot phase of the CGM. However, some fraction may contribute directly to the cool gas, either by accretion through streams penetrating the virial shock \citep{Mandelker17}, or infall of massive clouds that are not completely disrupted \citep{Afruni22}. 

Satellite galaxies can contribute some cool gas to the CGM. For example, the gas stripped or expelled from the LMC is responsible for the Magellanic Stream and (possibly) the Leading Arm. \cite{Putman21} show that the small dwarf galaxies inside the virial radius of the MW and M31 are all gas poor, while a few dwarf galaxies detected at larger distances have retained their gaseous components \citep{RW08,Gio13}. Presumably, at least some of the gas content of the accreted satellites was added to the MW cool CGM reservoir, and we expect the same to happen in other galaxies.

Cooling from the warm/hot phase can also contribute to the cool phase through condensation and cloud formation, although the transformation between gas phases in the CGM is highly uncertain. Addressing the conditions for condensation and precipitation, \cite{Sharma12a} found that warm/hot gas with a low ratio of cooling to dynamical time, $\tcool/\tdyn$ is more susceptible to condensation. This was adopted by \citet{Voit19} as the main parameter regulating gas accretion onto the galaxy in the precipitation model (however, see \citealp{Esmerian21}). In the \citetalias{FSM20} fiducial model, $\tcool/\tdyn$ decreases with distance from the galaxy, and $\tcool/\tdyn(r>50~{\rm kpc}) \sim 4$, below the threshold of $10$ estimated in \cite{McCourt12} and adopted in the precipitation model. This suggests formation of cool gas occurs at large distances from the galaxy, consistent with the extended cool gas distribution in the COS-Halos data, and in our model. After formation, the cool gas may migrate inwards (see Cruz~et~al., in~prep.). This is qualitatively different from the picture in \citet{MB04}, for example, where the cool gas forms at small radii first, where the absolute dynamical and cooling times are shorter (see also \citealp{Marinacci10} and \citealp{Fraternali17} for fountain- and accretion-induced condensation close to the galactic disk).

For example, in the \citetalias{FSM20} fiducial model, the global, mean mass cooling rate of the warm/hot CGM is $13.3$~\msuny. \citetalias{F22} calculate the cooling rate of the warm/hot CGM as a function of the CGM mass for a sample of galaxies generated by the Santa-Cruz semi-analytic model \citep{Somer99,Somer08,Somer15}. They get a similar result, with $\dot{M}_{\rm w/h} \approx 9.4 \times 10^{0.21 \sigma} \times \left(\mhot /5 \times 10^{10}~\msun \right)^{1.64}$~\msuny, where $\sigma$ accounts for the distribution in metallicity at a given CGM mass. While the \citetalias{FSM20} model assumes that the warm/hot gas is in equilibrium, and the cooling rate is balanced by heating from galactic feedback, the balance does not have to be perfect. Some fraction of the net mass cooling rate of the warm/hot CGM may accrete directly onto the galaxy \citep{Joung12b}, and the rest can replenish the cool CGM reservoir. 

Finally, galactic outflows are commonly observed in star forming galaxies in the low-redshift universe \citep{Rupke05a,Martin12,Bolatto13,Chisholm17a,Werk19,Rubin22} and can add cool gas to the CGM (see \citealp{Fraternali17,Veil20rev} for reviews)\footnote{~We do not discuss winds driven by accretion onto super-massive black holes (SMBHs) since the COS-Halos galaxies do not host Active Galactic Nuclei (AGN). While the effects of past AGN-driven outflows may be important, addressing these is beyond the scope of this work.}. Mass outflow rates are challenging to estimate observationally due to uncertainties in gas ionization fractions, metallicities, velocity distributions, and geometry. \cite{Chisholm17a} use UV observations to study outflows in a sample of nearby galaxies with $10^{7}<M_*/\msun<10^{11}$. For MW-mass galaxies, with $M_* \sim 5 \times 10^{10}$~\msun, they infer a mass loading factor of $\dot{M}/SFR \approx 0.3$ (see their Fig.~1). For $SFR\approx 4.3$~\msuny, the median of the COS-Halos sample, this corresponds to an outflow rate of $\sim 1.5$~\msuny, and suggests galactic winds can offset a significant fraction of the cool accretion rate. In the MW, \cite{Fox19} use metal absorption in HVCs to estimate $\dot{M} \approx 0.16 (Z'/0.5)$~\msuny, providing a lower limit to the total outflow rate.

Outflow rates are easier to calculate in hydrodynamical simulations, although they may depend on the feedback implementation. For $10^{12}$~\msun~halos at $z<0.5$, recent numerical studies of galaxy disks \citep{Kim20} and cosmological zoom-in simulations \citep{Christensen16,Tollet19,Pandya21} find $\dot{M}/SFR \sim 0.1-0.5$, similar to the observational estimates. \cite{Mitchell20} find values higher by factor of $2-3$ in EAGLE, suggesting outflows may be dominant in replenishing the cool CGM.

However, in non-starburst galaxies, the spatial extent of winds may be limited, and this channel is probably more significant close to the galaxy, at $r \lesssim 10-30$~kpc. As shown in \S\ref{subsec:mod_radius}, the measured columns are consistent with an extended distribution of cool gas, with $r_{\rm out} \gtrsim 0.6 \rvir \approx 150$~kpc, and possibly all the way out to the virial radius. This large spatial extent suggests that the build-up of the cool CGM reservoir took place at earlier times, when SFR were significantly higher, or through a different channel. Furthermore, the metallicity of $Z' \sim 0.3$ is lower than expected from enriched galactic outflows and than the metallicity of the warm/hot gas in the fiducial \citetalias{FSM20}. This may suggest that either (i) the cool phase is less dominated by outflows, or (ii) mixing of enriched outflowing gas with metal-poor inflows is less efficient for the cool phase than in the more diffuse warm/hot medium.

Related to this discussion, a series of recent studies explored different scenarios for the origin of the cool CGM in low-redshift galaxies \citep{Afruni19,Afruni21,Afruni22}. For example, \cite{Afruni21} model the kinematics of the cool gas in the COS-Halos galaxies with an outflow scenario. They find that the COS-Halos cool gas cannot be a result of SNe outflows, since the energy required for that would be more than available in SNe events. \cite{Afruni19} and \cite{Afruni22} perform similar modeling with quiescent galaxies (COS-LRG) and M31 (AMIGA) and obtain similar results, concluding that most of the cool gas comes from IGM accretion. In these works, most of the cool clouds are at large radii from the galaxies, resulting in a flat column density distribution with impact parameter (see Fig.~5 in \citealp{Afruni22}). We find that gas has to be distributed across a range of radii, from close to the galaxy and possibly out to \rvir, to reproduce the binned steep profiles of HI and the low metal ions (see Figures~\ref{fig:hi}-\ref{fig:silicon}).

\vspace{0.2 cm}

To summarize this section, we have shown that the reservoir of cool CGM can be replenished in several ways, either through accretion from the IGM (smooth or as dwarf galaxies), condensation from the warm/hot phase, or outflows from the galaxy. The estimated rates suggest that even if the cool CGM accretes onto the galaxy at a rate equal to the SFR, its net depletion rate can be close to zero, leading to a mean constant mass of cool gas in the CGM, and approximate equilibrium on long, possibly cosmological timescales. It is interesting to note that in the COS-Halos sample, lines of sight probing the CGM of quiescent galaxies show similar ionization states and column densities. Other observational studies of cool gas around quiescent galaxies at $z<1$ find similar results (\citealp{Zahedy19a,Chen20,Qu22}) This work does not aim to answer the question of what causes galaxies to stop forming stars and what part the (cool) CGM plays in it (see also \citealp{KT23}). Our model can be applied to observations of quiescent galaxies in a future study.

\section{Summary}
\label{sec:summary}
 
In this work we presented a model for the cool, photoionized phase of the CGM that includes density variation with radius, allows for non-thermal pressure, and addresses the gas thermal and ionization state. Applying the cool gas model to the COS-Halos absorption measurements of star-forming, MW-mass galaxies at low redshifts allows us to constrain the gas mass, volume filling fraction, amount of non-thermal support, and spatial distribution.

In \S\ref{sec:setup} we presented our model setup for cool gas in heating/cooling and photoionization equilibrium with the MGRF. We assumed that the cool gas in the CGM is in total pressure equilibrium with the warm/hot, volume filling phase, allowing it to be long-lived, consistent with the high detection rate measured in the COS-Halos survey. Our model allows for non-thermal pressure support that may be different from that in the warm/hot gas, leading to cool gas densities that are lower than expected from thermal pressure equilibrium (Figure~\ref{fig:thermal}), as suggested by \cite{Werk14}.

In \S\ref{sec:models} we applied our model to the star-forming galaxies in the COS-Halos data set. For the ambient warm/hot phase in which the cool gas resides we adopted the FSM20 fiducial model, reproducing the \OVI~observations of the same sample. We presented a nominal model and three parameter variations (see Table~\ref{tab:inputs}), in the amount of non-thermal support (\S\ref{subsec:mod_pressure}), gas spatial extent (\S\ref{subsec:mod_radius}), and the cool gas mass (\S\ref{subsec:mod_mass}). We show that our nominal model parameter set, with $\eta =3$ and $\mcool = 3 \times 10^9$~\msun, successfully reproduces the mean observed column densities of \HI~(Figure~\ref{fig:hi}) and the low/intermediate  metal ions - \CII, \CIII, \SiII, \SiIII, and \MgII~(Figures~\ref{fig:carbon}-\ref{fig:mg2col}). Variation of $\pm 0.5$~dex in the amount of non-thermal support and gas mass reproduces $\sim 2/3$ of the observations for the low/intermediate ions. The nominal model is consistent with the \SiIV~upper limits in the data set, but a factor of $\sim 3$ and $\sim 10$ lower than the \SiIV~and \CIV~measurements, respectively.

In \S\ref{sec:advanced} we discussed scenarios that reproduce the high observed $\HI$, \CIV, and \SiIV~column densities. First, we show that a model with $\mcool = 10^{10}$~\msun~and $\eta=1$ produces large amount of low-ionization gas, resulting in high \HI~columns, consistent with those observed in $\approx 20\%$ of the objects. Similarly, a model with $\mcool = 10^{10}$~\msun~and $\eta=5$, produces high \CIV~and \SiIV~columns, close to those measured in some objects (Figure~\ref{fig:adv_mod}). We also show that the high ions can, in principle, form in intermediate temperature gas, cooling from the hot phase or in residing in mixing layers around the cool clouds. However, our calculation shows this scenario requires a high volume filling fraction ($\sim 1/2$) and high mass ($\sim 1.5 \times 10^{10}~\msun)$, which may be in tension with recent numerical studies of this phase. We then presented a brief exploration of the model parameter space, and showed that for a volume filling fraction and metallicity that are constant with radius, using the columns of \HI~and two metal ions allows to estimate the gas metallicity, mean volume density, and total mass for a given object/sightline (Figures \ref{fig:adv_estim1}-\ref{fig:adv_estim2}).

In \S\ref{sec:uncertain} we addressed some of the model assumptions and uncertainties. First, we estimate how the assumed distribution of the ambient phase affects our modeling results by repeating our calculations with different warm/hot gas pressure profiles. Second, we address the uncertainty in the radiation field ionizing the cool CGM. The MGRF ionizing flux may be a factor of $1.5$ higher, reducing the amount of non-thermal support by the same factor. Our estimates suggest that for the typical SF galaxy, radiation from the galaxy or galactic outflows is probably sub-dominant beyond $\sim 15-30$~kpc, or $0.05-0.10$~\rvir.

While our model is agnostic (by construction) to the origin of the cool CGM and its small-scale morphology, our modeling results provide some constraints or clues about these. In \S\ref{subsec:disc_sizes} we show that the gas volume filling fraction, combined with the measured sky covering fraction of cool gas, gives and upper limit on the cloud sizes. We find that for the nominal model, $\rcl \lesssim 0.5$~kpc to be consistent with a covering fraction of $\sim 1$ and $\approx 3$ absorption components (see Figure~\ref{fig:morphology}). We also show that these clouds sizes are similar to the cloud sizes produced by thermal shattering for the gas densities in our models. Finally, in \S\ref{subsec:disc_res} we address the depletion and replenishment rates of the cool CGM reservoir. We estimate a mean accretion rate of $\sim 3$~\msuny~for our nominal model, similar to the median star formation rate in the COS-Halos galaxies. We then discuss possible channels for cool gas formation in the CGM and show that its mass can be in a state of approximate equilibrium, also consistent with the high detection rate of cool gas in observations.

\vspace{0.1cm}

In this work we aimed to reproduce the distribution of cool gas in the CGM of a typical SF galaxy at low redshift, as probed by the mean \HI~and metal ion column densities. To do this, we presented a simple yet flexible model, and showed that the inferred model parameters allow to constrain the underlying gas properties, such as density, spatial distribution, mass, and even cloud sizes. Our results show that (i) gas masses of $10^{9}-10^{10}$~\msun~in the cool phase, a fraction of the total CGM mass, and (ii) thermal pressure ratios that ranges from $1 < P_{\rm cool,th}/P_{\rm hot,th} < 10$ are consistent with the observations, and more extreme scenarios are not necessary.

This work is just the first step. First, as we noted earlier, modeling of individual sightlines, rather than the mean columns, will allow to further test the model, and may provide insights into the relation between galaxy and CGM properties. Second, in this work we assumed that the gas metallicity, volume filling fraction, and the non-thermal support are constant functions of radius. While this allows a better understanding of the model, in reality, the distributions of the properties are probably more complex, and different scenarios can be tested with our model framework in the future. Finally, applying this model to data probing larger distances, and to the CGM of galaxies residing in a wider range of halo masses and SF rates may allow us to test different scenarios for the formation of cool gas in the CGM, its interaction with the warm/hot gas, and the role the multiphase CGM plays in galaxy formation and evolution.

\begin{acknowledgements}

We thank Greg Bryan, Akaxia Cruz, Thomas Do, Drummond Fielding, Cassie Lockhhaas, Chris McKee, Matt McQuinn, Stella Ocker, Kartick Sarkar, Rachel Somerville, Amiel Sternberg, Kirill Tchernyshyov, and Fakhry Zahedy for helpful discussions during the course of this work. We thank Matt McQuinn for their questions, suggestions, and comments on the manuscript. YF thanks Chris McKee and Amiel Sternberg for collaboration on previous incarnations of this project, and for discussions during this work. YF thanks Liam Becker and Cyrus Taidi for the discussions held during the UW Pre-MAP project, and the CCA for hospitality during the TAU-CCA workshop. This research was supported by NASA award 19-ATP19-0023. JKW gratefully acknowledges support from the Research Corporation for Science Advancement under grant ID number 26842, and from the NSF-CAREER 2044303. This work also benefited from discussions held during the program "Fundamentals of Gaseous Halos" at the KITP (UCSB), supported by the National Science Foundation under Grant No.~NSF PHY-1748958.

\end{acknowledgements}


\renewcommand{\theequation}{A-\arabic{equation}}
 \renewcommand{\thefigure}{A-\arabic{figure}}
 \renewcommand{\thetable}{A-\arabic{table}}
 \setcounter{equation}{0}
 \setcounter{figure}{0}
 \setcounter{section}{0}
 \setcounter{table}{0}

\section*{Appendix~A - Summary of Observational Data}
\label{sec:app_obstab}

In Table~\ref{tab:obscol} we list the properties of the galaxies in the sample addressed in this study (redshift, stellar mass, and star formation rate), the sightlines through the CGM (impact parameter, physical and normalized to $\rvir$), and the hydrogen and metal column densities reported for them. The galaxy properties are taken from \cite{Werk12} (Table 4), and the absorption measurements are taken from \citetalias{Werk13} (Table~4) and \citetalias{Prochaska17} (Tables~2 and 3). All the data are taken as is and provided for ease of comparison to the models presented in this work.


\bgroup
 \def\arraystretch{2.0}

\begin{table*}
	\centering
	\rotatebox{90}{
		\begin{minipage}{\textheight}
 	\caption{Observational Data Summary} 
 	
 \begin{tabular}{| l | c || c | c | c || c | c | c | c | c | c | c | c | c | c |}

 \midrule
 SDSS Field & Galaxy ID & $z$ & ${\rm log}(M^*/\msun)$ & SFR ($\msuny$) & $h$ (kpc) & $h/\rvir$ & \HI & \CII & \CIII & \CIV & \SiII & \SiIII & \SiIV & \MgII\\
 \midrule
 J0401-0540 & 67\_24 & $0.2197$ & $10.14$ & $1.14 \pm 0.15$ & $83$ & $0.35$ & $15.39_{-0.05}^{+0.06}$ & $<13.58$ & $>14.00$ & --- & $<12.54$& $12.88 \pm 0.06$ & $<13.07$ & $<12.26$ \\
 J0910+1014 & 34\_46 & $0.1427$ & $10.61$ & $14.42 \pm 1.64$ & $112$ & $0.34$ & $17.71_{-0.26}^{+0.23}$ & $14.16 \pm 0.06$ & --- & $14.10 \pm 0.09$ & $12.96 \pm 0.08$ & $>13.28$ & --- & $>13.29$ \\
 J0914+2823 & 41\_27 & $0.2443$ & $9.81$ & $2.83 \pm 0.34$ & $101$ & $0.51$   & $15.40_{-0.06}^{+0.05}$ & $<13.64$ & --- & --- & $<12.63$ & $12.63 \pm 0.11$ & $<13.12$ & $<11.87$ \\
 
 J0943+0531 & 227\_19 & $0.3530$ & $9.59$ & $0.47 \pm 0.11$ & $92$ & $0.55$   & $16.28 \pm 0.04$        & $14.42 \pm 0.09$ & $>14.30$ & --- & $<13.25$ & $<12.97$ & --- & $<12.44$ \\
 J0943+0531 & 106\_34 & $0.2284$ & $10.79$ & $4.52 \pm 0.58$ & $121$ & $0.34$   & $15.94_{-0.41}^{+0.45}$       & $<13.67$ & --- & --- & $<12.70$     & $12.89 \pm 0.10$ & $<13.40$ & $<12.20$ \\
 J1009+0713 & 204\_17 & $0.2278$ & $9.85$ & $4.58 \pm 0.61$ & $60$ & $0.29$ & $17.26_{-0.13}^{+0.13}$ & $14.37 \pm 0.05$ & --- & --- & $<13.16$ & $>13.26$ & $<13.44$ & $12.68 \pm 0.03$ \\
                
J1016+4706 & 274\_6 & $0.2520$ & $10.21$ & $0.53 \pm 0.06$ & $23$ & $0.10$ & $17.05_{-0.05}^{+0.05}$ & $14.68 \pm 0.03$ & $>14.58$ & ---& $13.81 \pm 0.05$ & $>13.87$ & --- & $>13.55$ \\
J1016+4706 & 359\_16 & $0.1661$ & $10.48$ & $1.37 \pm 0.17$ & $44$ & $0.15$   & $17.74_{-0.94}^{+0.50}$ & $14.27 \pm 0.07$ & --- & ---     & $13.68 \pm 0.06$ & $>13.74$ & --- & $>13.34$ \\
J1112+3539 & 236\_14 & $0.2467$ & $10.31$ & $5.68 \pm 0.80$ & $53$ & $0.21$   & $16.68_{-0.48}^{+0.52}$ & $<13.91$ & --- & ---             & $<12.93$ & $12.09 \pm 0.11$ & $>13.38$ & $12.38 \pm 0.09$ \\

J1133+0327 & 164\_21 & $0.1545$ & $10.08$ & $1.83 \pm 0.22$ & $55$ & $0.27$   & $>15.09$                & $<13.77$ & --- & ---             & $<13.07$ & $<12.91$ & --- & $<13.07$ \\
J1233-0031 & 168\_7 & $0.3185$ & $10.53$ & $3.42 \pm 0.54$ & $32$ & $0.12$   & $15.52_{-0.10}^{+0.07}$ & $<13.65$ & $>14.18$ & ---        & $<13.33$ & $13.01 \pm 0.10$ & --- & $<12.39$ \\
J1233+4758 & 94\_38 & $0.2221$ & $10.76$ & $4.38 \pm 0.52$ & $132$ & $0.38$   & $16.74_{-0.06}^{+0.06}$ & $>14.49$ & --- & ---             & $13.45 \pm 0.05$ & $>13.44$ & $13.38 \pm 0.07$ & $>13.37$ \\

J1241+5721 & 208\_27 & $0.2178$ & $10.04$ & $1.06 \pm 0.17$ & $93$ & $0.41$   & $15.21_{-0.12}^{+0.08}$ & $<13.43$ & --- & ---             & $12.84 \pm 0.10$ & $12.72 \pm 0.09$ & $13.28 \pm 0.11$ & $<12.32$ \\
J1245+3356 & 236\_36 & $0.1925$ & $9.84$ & $1.05 \pm 0.17$ & $112$ & $0.54$   & $14.66_{-0.06}^{+0.06}$ & $<13.57$ & $13.52 \pm 0.07$ & ---  & $<13.19$ & $<12.33$ & $<12.84$ & $<11.92$ \\
J1330+2813 & 289\_28 & $0.1924$ & $10.32$ & $1.99 \pm 0.23$ & $87$ & $0.33$   & $17.01_{-0.13}^{+0.11}$ & $14.37 \pm 0.05$ & $>14.08$ & --- & $13.49 \pm 0.04$ & $13.19 \pm 0.04$ & $<13.29$ & $>13.36$ \\
J1342-0053 & 157\_10 & $0.2270$ & $10.93$ & $6.04 \pm 0.74$ & $35$ & $0.09$   & $18.82_{-0.15}^{+0.07}$  & $>15.13$ & $>14.65$ & ---          & $>14.49$ & $>13.98$ & $13.49 \pm 0.11$ & $>14.08$ \\
J1419+4207 & 132\_30 & $0.1792$ & $10.61$ & $3.77 \pm 1.06$ & $88$ & $0.28$   & $16.89_{-0.20}^{+0.18}$ & $14.06 \pm 0.08$ & $1>14.11$ & --- & $13.28 \pm 0.07$ & $>13.29$ & $<13.53$ & $>13.21$ \\

J1435+3604 & 68\_12 & $0.2024$ & $11.09$ & $18.96 \pm 2.28$ & $39$ & $0.08$   & $19.73_{-0.14}^{+0.09}$ & $>14.61$ & $>14.31$ & ---          & $>14.24$ & $>13.46$ & $<13.29$ & $>13.74$ \\
J1435+3604 & 126\_21 & $0.2623$ & $10.37$ & $5.56 \pm 0.70$ & $83$ & $0.32$   & $15.17_{-0.10}^{+0.11}$ & --- & --- & ---                    & $<12.77$ & $12.84 \pm 0.10$ & $<13.46$ & $<12.21$ \\
J1437+5045 & 317\_38 & $0.2460$ & $10.14$ & $4.29 \pm 0.50$ & $143$ & $0.73$   & $14.53 \pm 0.12$        & --- & --- & ---                    & $<13.04$ & $<12.82$ & $<13.38$ & $<12.22$ \\
 
J1445+3428 & 232\_33 & $0.2176$ & $10.40$ & $2.60 \pm 0.31$ & $113$ & $0.39$   & $15.07 \pm 0.06$        & $<13.62$ & --- & ---               & $<13.01$ & $<12.63$ & $<13.21$ & $<12.55$ \\
J1550+4001 & 97\_33 & $0.3218$ & $10.90$ & $7.41 \pm 0.96$ & $150$ & $0.48$   & $13.86 \pm 0.09$        & $<13.98$ & $<12.96$ & ---          & $<12.96$ & $<12.71$ & --- & $<11.53$ \\
J1555+3628 & 88\_11 & $0.1893$ & $10.53$ & $3.77 \pm 1.06$ & $34$ & $0.11$   & $17.31_{-0.14}^{+0.15}$ & $14.49 \pm 0.05$ & $>14.30$ & ---  & $13.41 \pm 0.05$ & $>13.53$ & $13.88 \pm 0.12$ & $13.26 \pm 0.05$ \\
J1619+3342 & 113\_40 & $ 0.1414$ & $10.11$ & $1.33 \pm 0.17$ & $97$ & $0.39$   & $14.96 \pm 0.03$        & $14.30 \pm 0.04$ & --- & $13.90 \pm 0.03$            & $<12.42$ & --- & $13.19 \pm 0.08$ & $<12.23$ \\
\bottomrule
 	\end{tabular}
 	  \label{tab:obscol}
 	\end{minipage}
 	}
 	\end{table*}
 	
 	  	  

\renewcommand{\theequation}{B-\arabic{equation}}
 \renewcommand{\thefigure}{B-\arabic{figure}}
  \renewcommand{\thetable}{B-\arabic{table}}
 \setcounter{equation}{0}
 \setcounter{figure}{0}
 \setcounter{section}{0}
 \setcounter{table}{0}

\section*{Appendix~B - Radial Distributions of Metal Ions}
\label{sec:app_spatial}

In \S\ref{sec:models} we presented the column density profiles for model with different parameter combinations, and compared them to observations. We now present and discuss the underlying radial profiles of the ion fraction and ion volume density for carbon. As discussed in Appendix~C, the neutral hydrogen fraction is approximately proportional to the gas density, and the silicon ions are similar in their behavior to the carbon ions. The radial fraction and density profiles for all the ions discussed in the paper are available in the data files are attached to this manuscript.

We focus on the models in the Pressure set ($\#1$), which vary in the value of $\eta$, the amount of non-thermal pressure in the cool gas, affecting the gas density and the ion fractions. The profiles for the models in the other sets can be calculated from the profiles of the nominal model, and we address this at the end of this section.

The top panels of Figure~\ref{fig:zap2_all} show the ion fractions of \CII~(left panel), \CIII~(middle), and \CIV~(right). In the nominal model (solid curve), the \CII~fraction decreases rapidly with distance from the halo center, as the gas density decreases and it becomes more ionized. The fraction of \CIV~shows an opposite trend, increasing strongly with $r$. The \CIII~ fraction decreases with radius up to $r \sim 0.5\rvir$ and decreases at larger radii, and it is higher than $0.1$ almost everywhere in the halo. The low-$\eta$ model (dashed curves) has higher gas volume densities (see middle panel of Figure~\ref{fig:thermal}), leading to less ionized gas compared to the nominal model. As a result, the \CII~(\CIV) fraction is higher (lower), by about an order of magnitude at $r \sim 0.6 \rvir$. The high-$\eta$ model (dotted curves) has lower gas densities, resulting in lower (higher) \CII~(\CIV). In all three models, the \CII~fraction peaks close to the halo center, where gas density is highest, with $f_{\CII} \sim 1$. For \CIV, the fraction is highest close to or at the outer boundary, and does not go above $f_{\CIV} \sim 0.3$, the maximal $\CIV$ fraction in photoionized gas.

\begin{figure*}[t]
\includegraphics[width=0.99\textwidth]{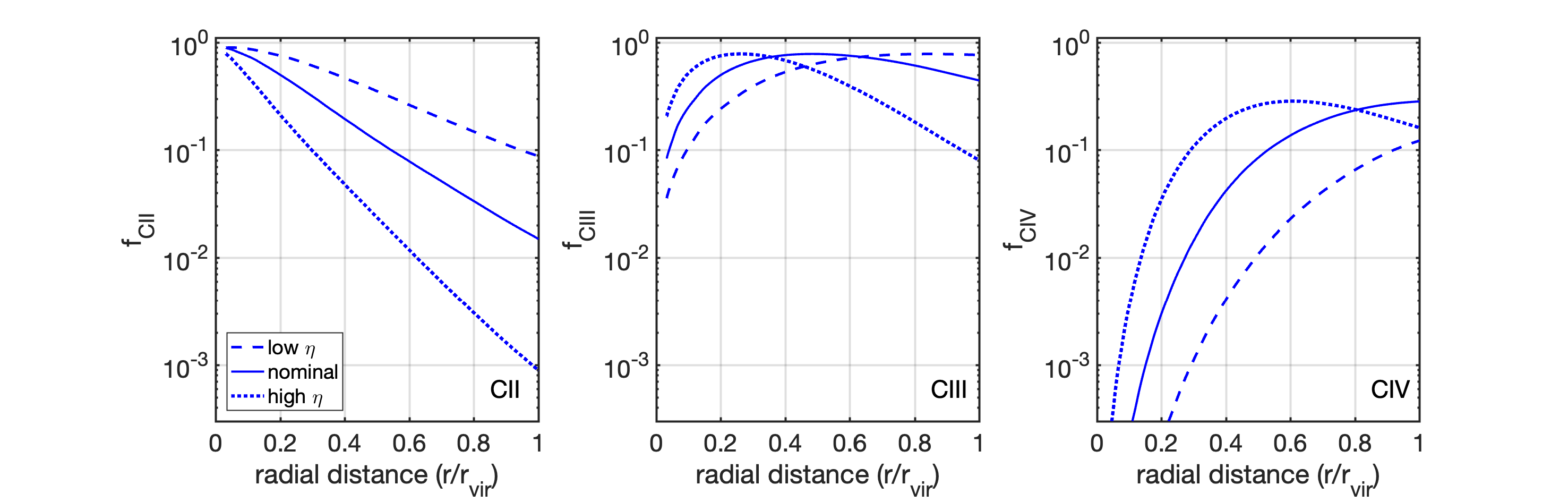}
\includegraphics[width=0.99\textwidth]{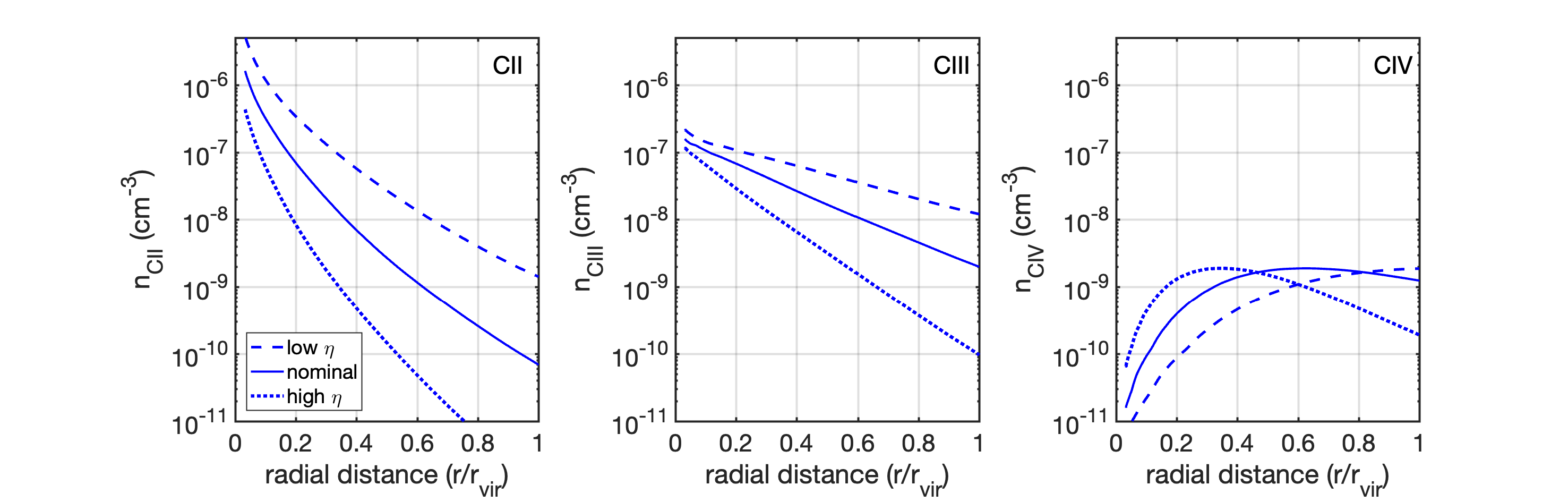}
 \caption{Radial profiles of the carbon ion fractions (top) and densities (bottom) - \CII~to \CIV~(left to right) for the models in the Pressure set ($\#1$, see \S\ref{subsec:mod_pressure}).}
  \label{fig:zap2_all}
\end{figure*}

The bottom panels of Figure~\ref{fig:zap2_all} shows the ion volume densities, given by the product of the ion fractions (shown in the top panels), solar carbon abundance, $A_C = 2.7 \times 10^{-4}$, model metallicity, $Z' = 0.3$~solar, and the hydrogen density profiles, plotted in the middle panel of Figure~\ref{fig:thermal}.

The \CII~densities decrease rapidly with radius due to a combination of the steep gas density and ion fraction radial profiles, resulting in steep volume density profiles and steep column density profiles as function of impact parameter (see top panels of Figure~\ref{fig:carbon}}). The low- (high-) $\eta$ model has higher (lower) densities and \CII~fractions, leading to \CII~densities higher (lower) by $0.5-1.0$~dex than in the nominal model at all radii. The shape of the \CIII~fraction profile leads to a weak variation in the volume density, of $0.5-1.0$ dex, across the full radial range. The difference between models is also smaller here, with a maximum of $\sim 0.5$ dex at $r \sim 0.5$~\rvir. Finally, the \CIV~fraction increases more rapidly with radius than the gas density decreases, resulting in much smaller variation in the volume density. The opposite trends of the gas density and ion fraction lead to the \CIV~being lower (higher) in the low- (high-) $\eta$ model compared to the nominal model. The steep profiles can be approximated by a thick shell distribution, and result in close to flat column density profiles for \CIV~(bottom panels of Figure~\ref{fig:carbon}).

Finally, the ion fractions and volume densities presented can be used useful to easily calculate the profiles for additional models, including those in model sets $\#2$ and $\#3$. First, the ion fractions are not affected by the gas radial extent, $r_2$, or the gas volume filling fraction, $\fvc$, and their profiles are identical for model with the same density profile and different combinations of $r_2$ and $\fvc$. For the models in the Radius and Mass sets ($\#2$ and $\#3$), the ion fraction profiles are identical to those of the nominal model presented here. Second, the ion volume densities in our models are proportional to $\fvc$, and can be calculated by scaling the profiles of the nominal model by the ratio of the volume filling factors, given in Table~\ref{tab:inputs}.


\addcontentsline{toc}{section}{References}

\bibliographystyle{yahapj}


\end{document}